\documentclass[12pt]{article}
\usepackage{authblk}
\renewcommand\footnotemark{}

\usepackage{amsmath}
\usepackage{graphicx}
\usepackage{enumerate}
\usepackage{natbib}
\usepackage{url} 

\newcommand{\blind}{1}

\usepackage{amsmath,amsfonts}
\usepackage{natbib}
\usepackage{url}  
\usepackage{bm}
\usepackage{xcolor}
\usepackage{array}
\usepackage{caption}

\usepackage{comment}

\newcommand{\bx}{\bm x}

\newcommand{\bmeta}{\bm \eta}

\newcommand{\bomega}{\bm \omega}
\newcommand{\blambda}{\bm \lambda}
\newcommand{\balpha}{\bm \alpha}
\newcommand{\btheta}{\bm \theta}
\newcommand{\bphi}{\bm \phi}
\newcommand{\bgamma}{\bm \gamma}

\newcommand{\bdelta}{\bm \delta}

\newcommand{\bpi}{\bm \pi}
\newcommand{\bV}{\bm V}

\newcommand{\TT}{\mathcal{T}}
\newcommand{\UU}{\mathcal{U}}
\newcommand{\LL}{\mathcal{L}}

\newcommand{\bOne}{\bm{1}}
\newcommand{\nameholder}{LCVA}

\usepackage{tikz}
\usepackage{environ}
\makeatletter
\newsavebox{\measure@tikzpicture}
\NewEnviron{scaletikzpicturetowidth}[1]{%
  \def\tikz@width{#1}%
  \begin{lrbox}{\measure@tikzpicture}%
  \BODY
  \end{lrbox}%
  \pgfmathparse{#1/\wd\measure@tikzpicture}%
  \BODY
}
\usetikzlibrary{fit,positioning}

\usepackage[rightbars, color]{changebar}
\newcommand{\rev}[1]{%
  \cbcolor{white}
  \begin{changebar}
    #1
  \end{changebar}%
  }%

\begin{document}

\def\spacingset#1{\renewcommand{\baselinestretch}%
{#1}\small\normalsize} \spacingset{1}


\if1\blind
{
  \title{\bf Bayesian Nested Latent Class Models for Cause-of-Death Assignment using Verbal Autopsies Across Multiple Domains}
  \author[1]{Zehang Richard Li\thanks{
    ZRL was supported by grant R03HD110962 from the Eunice Kennedy Shriver National Institute of Child Health and Human Development (NICHD). ZW and IC were supported in part by a seed grant from Michigan Institute of Data Science. SJC was supported by grant R01HD086227 from the Eunice Kennedy Shriver National Institute of Child Health and Human Development (NICHD).}}
  \author[2]{Zhenke Wu}
  \author[2]{Irena Chen}
  \author[3]{Samuel J. Clark}
  \affil[1]{Department of Statistics, University of California, Santa Cruz}
  \affil[2]{Department of Biostatistics, University of Michigan}
  \affil[3]{Department of Sociology, The Ohio State University}
  \maketitle
} \fi

\if0\blind
{
  \bigskip
  \bigskip
  \bigskip
  \begin{center}
    {\LARGE\bf Bayesian Nested Latent Class Models for Cause-of-Death Assignment using Verbal Autopsies Across Multiple Domains}
\end{center}
  \medskip
} \fi

\bigskip
\begin{abstract}
Understanding cause-specific mortality rates is crucial for monitoring population health and designing public health interventions. Worldwide, two-thirds of deaths do not have a cause assigned. Verbal autopsy (VA) is a well-established tool to collect information describing deaths outside of hospitals by conducting surveys to caregivers of a deceased person. It is routinely implemented in many low- and middle-income countries. Statistical algorithms to assign cause of death using VAs are typically vulnerable to the distribution shift between the data used to train the model and the target population. This presents a major challenge for analyzing VAs as labeled data are usually unavailable in the target population. This article proposes a Latent Class model framework for VA data (LCVA) that jointly models VAs collected over multiple heterogeneous domains, assigns causes of death for out-of-domain observations, and estimates cause-specific mortality fractions for a new domain. We introduce a parsimonious representation of the joint distribution of the collected symptoms using nested latent class models and develop a computationally efficient algorithm for posterior inference. We demonstrate that LCVA outperforms existing methods in predictive performance and scalability. Supplementary materials for this article and the R package to implement the model are available online.

\end{abstract}

\noindent%
\vfill

\spacingset{1.9} 

\section{Introduction}
\label{sec:intro}

Data describing cause of death is an essential component for understanding the burden of disease, emerging health needs, and effect of public health interventions. 
Few low- and middle-income countries (LMIC) have systems that produce high quality cause of death statistics. 
Only about two-thirds of the deaths worldwide are registered and up to half of these deaths are either not assigned a cause or assigned only an ill-defined cause \citep{world2021civil}. 
In many populations not served by official medical certification of causes, a technique known as verbal autopsy (VA) has been routinely used to infer causes of death.  
VA is a survey-based method whereby a structured questionnaire is conducted to a family member or a caregiver of a recently deceased person after a suitable mourning period. The VA interview collects information about the circumstances, signs, and symptoms leading up to the death. It is routinely used by researchers in Health and Demographic Surveillance System (HDSS) including the INDEPTH network \citep{Sankoh2012} and the ALPHA network \citep{maher2010translating}, 
multi-country research projects \citep{nkengasong2020improving,breiman2021postmortem}, 
and national scale surveys in many LMICs. 

Data collected by VAs are analyzed either by a panel of physicians or with statistical algorithms. The physician review approach can be effective if resource permits \citep{lozano2011performance}, but it is expensive and rarely possible for timely monitoring of cause-specific mortality. On the other hand,  statistical methods have been widely used to automate the cause-of-death assignment from VA data~\citep[e.g.,][]{byass2003probabilistic,miasnikof2015naive,serina2015improving,mccormick2016probabilistic}. These methods aim to construct a high-dimensional classifier for cause-of-death assignment and obtain population-level cause-specific mortality fraction (CSMF) estimates, i.e., the distribution of deaths due to each possible cause.

There are many  methodological challenges for cause-of-death assignment algorithms to be successfully deployed in practice.
First, all existing VA algorithms assume that there is a single training dataset or domain knowledge base from which the relationship between symptoms and causes can be learned.
In most situations where VA is used, local deaths from the target population with known causes are extremely rare. High-quality VA data with reference causes are usually only available outside of the target population, and may consist of data collected from several distinct study populations. 
This leads to the \emph{data shift} problem, where the source and target domain data may have different marginal and conditional distributions \citep{moreno2012unifying}, and it could significantly bias the cause-of-death assignment and CSMF quantification.
The issue of data shift received little discussion in the midst of developments of automated VA cause-of-death assignment algorithms in the last decade. 
The recent work of \citet{datta2020regularized} and \citet{fiksel2020generalized} are the first to address this issue by calibrating model predictions to additional local validation data with known causes of death. 
When no validation data exist, however, all existing VA methods are vulnerable to data shift. To our knowledge, there is no literature that characterizes the heterogeneous joint distributions of symptoms and causes across different domains or leverages such information for cause-of-death assignment.

Second, existing VA methods used by practitioners typically assume that symptoms are conditionally independent given the underlying cause of death as it reduces model complexity and computation cost significantly~\citep{byass2003probabilistic,miasnikof2015naive,mccormick2016probabilistic}. More recently, several approaches have been proposed to account for symptom dependence using latent Gaussian models with sparse or low-rank representations~\citep{li2020using,tsuyoshi2017,moran2021bayesian}. It has been shown that incorporating symptom dependence usually improves the accuracy of both individual cause-of-death assignment and the population-level CSMF estimation. However, the inferred dependence relationship among the continuous latent variables is difficult to interpret, as they do not correspond to the dependence of the observed binary symptoms \citep{li2020using}. In addition, the computation required to estimate the high-dimensional latent Gaussian mixtures can take hours or days to complete, which makes these methods difficult for routine use in low-resource settings.

In this paper, we address these challenges by developing a novel Latent Class model framework for VA data (\nameholder{}). 
A key feature of \nameholder{} is that we estimate a common collection of latent symptom profiles across different populations.
This leads to an explicit domain adaptation strategy for classifying out-of-domain deaths that takes into account the similarity between the target population and the existing training datasets. 
We introduce a sparse parameterization to further reduce the dimensionality of the latent symptom profiles that also facilitates direct interpretations of the conditional independence relationship among the observed symptoms.
Furthermore, our inference procedure is an order of magnitude faster than the existing methods using latent Gaussian models to capture symptom dependence. 

The rest of the paper is organized as follows. Section \ref{sec:transfer} reviews the background of existing VA methods and their implications under data shift. Section \ref{sec:latentclass} proposes the nested latent class model approach for cause-of-death assignment. Section \ref{sec:single-model} discusses the model with one training dataset. Section \ref{sec:multi-model} extends the model to the scenario where training data consist of deaths from multiple domains. Section \ref{sec:sampling} develops an efficient Markov chain Monte Carlo (MCMC) algorithm for posterior sampling. 
Section \ref{sec:results-phmrc} demonstrates the proposed model under various realistic scenarios using the Population Health Metrics Research Consortium (PHMRC) gold-standard VA dataset. Section \ref{sec:discussion} concludes the paper and discusses future directions.








\section{Domain adaption in cause-of-death assignment}
\label{sec:transfer} 
The domain adaptation problem refers to the situation where a statistical learning model trained on one labeled dataset needs to be generalized to the target dataset, or target domain, drawn from a different distribution and with insufficient labeled data \citep{daume2006domain}. 
In the context of cause-of-death assignment using VA, the discussion of cross-domain generalizability goes back to the early work of \citet{king2008verbal}, where the authors developed a constrained least square model trained on hospital deaths with known causes and subsequently applied to classifying community deaths. 
Let $X_i$ denote the vector of symptoms for individual $i$ and $Y_i$ denote the corresponding cause of death. For simplicity, let us assume for now that there are two datasets: a training dataset $\TT$ with both $X_i$ and $Y_i$ known, and a target dataset $\TT_0$ where only $X_i$ is observed. 
\citet{king2008verbal} makes the assumption that $p_{\TT_0}(X|Y) = p_{\TT}(X|Y)$. This assumption implies that the difference in the joint distribution of $p(X, Y)$ between the two domains can be fully explained by the difference in $p(Y)$, since $p(X, Y) = p(Y)p(X|Y)$. 
Prediction tasks under this decomposition are also termed anticausal prediction \citep{anticausal} and this type of data shift with domain-specific $p(Y)$ and domain-invariant $p(X|Y)$ is known as \emph{label shift} or \emph{prior shift} \citep{storkey2009training}.

A large collection of VA algorithms, however, took a more generic classification perspective. They first learn a model $p_{\TT}(Y|X)$ with the training data and then apply the classifier to observations from the target domain for cause-of-death assignment. 
The generalizability of this approach relies on the assumption that $p_{\TT_0}(Y|X) = p_{\TT}(Y|X)$. That is, the difference in the joint distribution of $p(X, Y)$ across domains are fully explained by different symptom distribution $p(X)$. This type of data shift is also known as \emph{covariate shift} \citep{shimodaira2000improving}. Under this assumption, 
the target CSMF can be estimated by simple aggregation of individual classifications, i.e.,
\begin{equation*}
  \hat p_{\TT_0}(Y) 
  = \int_x \hat p_{\TT_0}(Y | X = x) p_{\TT_0}(x)dx
  =  \int_x \hat p_{\TT}(Y | X = x) p_{\TT_0}(x)dx
\approx \frac{1}{n}\sum_{i=1}^n\hat p_{\TT}(Y | X_i).
  \label{eq:covariate-shift}
\end{equation*}
Algorithms that make the implicit assumption on covariate shift include some of the most widely adopted VA models, such as InterVA \citep{byass2012strengthening,byass2019integrated}, Tariff \citep{serina2015improving}, Naive Bayes Classifier \citep{miasnikof2015naive}, as well as many others in the literature applying off-the-shelf classification models on VA data \citep[e.g.,][]{flaxman2011random,blanco2020extracting}. 
Models assuming covariate shift are easy to design and implement, but difficult to justify in the context of VA. As pointed out by \citet{king2008verbal}, symptoms collected by VA are usually the consequences of the underlying causes of death. Thus the covariate shift assumption is almost always violated as $p_{\TT_0}(Y|X)$ and $p_{\TT}(Y|X)$
 are different even when $p(X|Y)$ is the same across domains.

The recent advancement of Bayesian modeling of VA data again returned to the formulation using the more natural decomposition of $p(X, Y) = p(Y)p(X|Y)$ to describe the data generating process.
The first Bayesian hierarchical model for cause-of-death assignment, InSilicoVA \citep{mccormick2016probabilistic}, proposed to use non-informative priors to model $p_{\TT_0}(Y)$ in the target population and strong informative priors to shrink $p_{\TT_0}(X|Y)$ to the observed $p_{\TT}(X|Y)$ calculated from training data or provided by domain experts.
This framework was later extended to include more flexible characterizations of $p(X | Y)$, using latent Gaussian graphical models \citep{li2020using} and factor models \citep{tsuyoshi2017, moran2021bayesian}. While \citet{tsuyoshi2017} and \citet{moran2021bayesian} treated the cause of death of all observations as drawing from the same distribution, they can be readily adapted to have different $p(Y)$ in different populations. 

\begin{figure}[!htb]
\centering
\includegraphics[width = \textwidth]{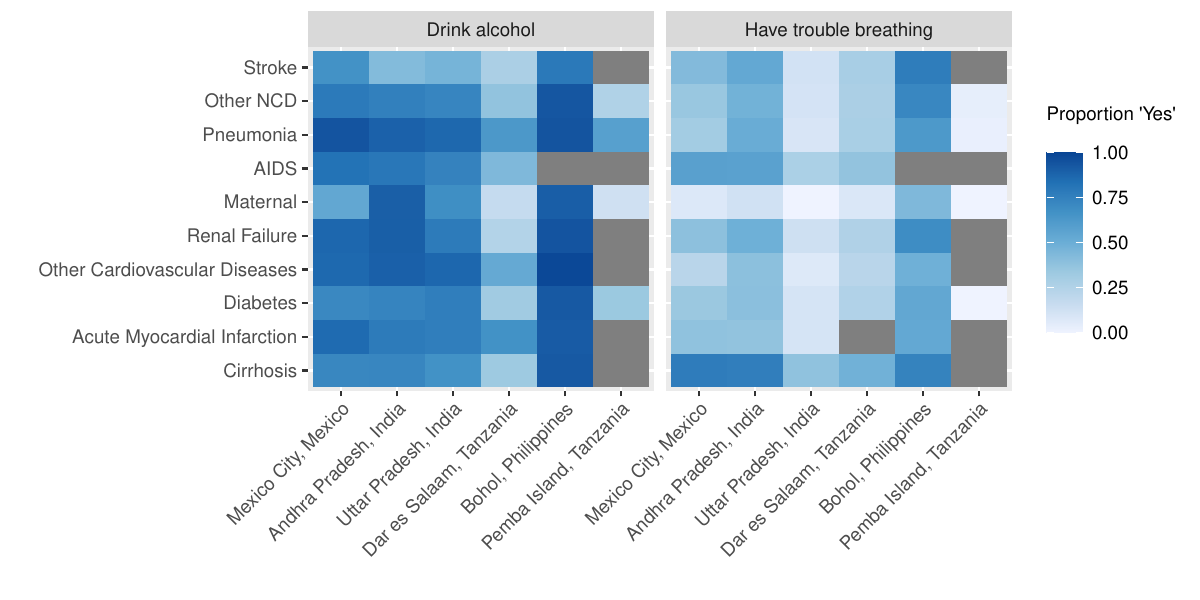}
\caption{Empirical proportion of `yes' response in two symptoms, drinking alcohol (left), and having trouble breathing (right), among deaths due to different causes of death in the six study sites in the PHMRC gold-standard dataset. Gray color indicates no collected death due to a given cause in a given site.}
\label{fig:illustration}
\end{figure}

The prior shift assumption, however, can still be overly simplistic in practice. 
Figure \ref{fig:illustration} illustrates an example from the PHMRC gold-standard VA dataset \citep{murray2011population} discussed in Section \ref{sec:results-phmrc}.
It shows the proportion of deaths with a `yes' response in two symptoms, \textit{having trouble breathing} and \textit{drinking alcohol}, conditional on the ten causes of death with most observations, for each of the six study sites in the dataset. 
\rev{
While it only illustrates the conditional distribution of two symptoms, strong domain heterogeneity can be observed from these site-specific empirical response probabilities $p(X|Y)$. 
Some of these reporting patterns may be explained by the presence of confounding factors \citep{moran2021bayesian}. However, we usually do not have enough data to identify or adjust for all potential confounders, many of which may be unobservable, or quantify the similarity of these distributions without reference cause of death.
When $p(X|Y)$ vary across domains, the choice of training data could have strong implications for out-of-domain predictions.}

A line of research closely related to this paper is the calibration models developed in \citet{datta2020regularized} and \citet{fiksel2020generalized}.
The focus of these methods are fundamentally different from our work. 
They consider a different scenario where the cause of death is known for a small sample of observations $\TT_0'$ from the target population, 
and assume that $p_{\TT_0'}(X | Y) = p_{\TT_0}(X | Y)$, i.e., only prior shift may exist between the labeled and unlabeled deaths from the same population, with potentially different $p_{\TT_0'}(Y)$ and $p_{\TT_0}(Y)$. 
It can be shown that this assumption is sufficient to de-bias a classifier trained on $\TT$ and applied to $\TT_0$, using the labeled target population data $\TT_0'$.
The advantage of such calibration methods is that they circumvent the need to identify and parameterize transportable component of the joint distribution from the training domains to the target domain. The cost, however, is the requirement for additional local data with known causes, which is usually impractical in resource-constrained settings. 
Moreover, the estimation of the error transition matrix of the predicted cause of death
is challenging when there are many potential causes of death. For a small number of local labeled data, \citet{datta2020regularized} and \citet{fiksel2020generalized} recommend grouping the causes of death into a handful of broad categories in the analysis. It remains an open problem to estimate the CSMF for longer cause lists with limited calibration data. 



The focus of this paper is to develop a general framework to characterize domain heterogeneity and leverage such information when assigning causes of death for out-of-domain observations, without additional calibration data. 
Unlike any existing VA models, we allow both $p(Y)$ and $p(X | Y)$ to vary across domains. 
The key to our modeling approach is a latent class model representation \citep{goodman1974exploratory} that assumes \emph{source component shift} for data across different domains \citep{storkey2009training}. That is, we assume that there is a latent variable $Z$ that renders $p_{\TT}(X | Y, Z) = p_{\TT_0}(X | Y, Z)$. This assumption can be considered as an extension of {prior shift} on the latent space. It allows us to combine data from multiple domains with domain-specific $p(X, Y)$ efficiently when assigning cause of death for unlabeled deaths in a new domain. We discuss our proposed model in more details in the next section.


\section{Nested latent class models for VA data}
\label{sec:latentclass}
In this section, we present the nested latent class model for VA data, \nameholder{}. We first introduce a single-domain latent class model that parsimoniously models the joint distribution of symptoms and causes, and then extend it to modeling data from multiple domains.   

\subsection{The single-domain model}
\label{sec:single-model}                                     
Let $Y_i \in \{1, ..., C\}$ denote the categorical cause of death for the $i$-th death and $X_i \in \{0, 1\}^p$ denote the $p$-dimensional reported binary symptom vector. The standard practice of VA usually involves pre-processing any continuous variables into binary indicators \citep{li2020using}. Thus we focus our discussion on binary symptoms here, but the model can easily incorporate categorical symptoms of more than two levels, thus mitigating the potential loss of information in the pre-processing step. We begin by assuming that each cause of death can be further divided into $K$ sub-categories and each individual is associated with a latent class $Z_i \in \{1, ..., K\}$. 
We assume reported symptoms are independent draws from a Bernoulli distribution given both the cause of death $Y_i$ and the latent class membership $Z_i$, i.e., 
\[
 X_{ij} | Y_i = c, Z_i = k \sim \mbox{Bern}(\theta_{ckj}), \;\; j = 1, ..., p.
\] 
We treat the cause of death $Y = (Y_1, ..., Y_n)$ and latent class indicator $Z = (Z_1, ..., Z_n)$ as random variables and let, for $i = 1, ..., n$,
\begin{align*}
    Y_i   &\sim \mbox{Cat}(\bpi),  \\
    Z_i | Y_i = c  &\sim \mbox{Cat}(\blambda_{c}).
\end{align*}
For deaths due to each cause of death, $p(X_i | Y_i = c) = \sum_{k=1}^K \lambda_{ck}\prod_{j = 1}^p \theta_{ckj}^{x_{ij}}(1-\theta_{ckj})^{1-x_{ij}}$. 
Conditional dependence among symptoms are introduced by marginalizing out the latent class indicators. With sufficiently many latent classes, this representation is flexible enough to represent any multivariate discrete distribution \citep{dunson2009nonparametric}. We treat the latent class $Z$ as defined within each cause of death and both latent indicators $Z$ and $Y$ need to be estimated for a new death. This nested formulation allows the model to be sufficiently expressive for deaths from each cause. 

The naive factorization of the joint distribution of $(X, Y, Z)$ requires $CK(p + 1) - 1$ parameters. This can be computationally challenging for VA data, as the number of symptoms $p$ and the size of cause list $C$ are usually large while sample size is typically small to moderate.
We adopt a sparsity-inducing prior to model the response probabilities $\theta_{ckj}$ proposed by \citet{zhou2015bayesian}. We let 
\begin{align*}
   \theta_{ckj} &= \delta_{ckj}\phi_{ckj} + (1 - \delta_{ckj})\gamma_{cj},
   \;\;\;\;
   \delta_{ckj} \sim \mbox{Bern}(\tau_c), \\
    \phi_{ckj} &\sim \mbox{Beta}(1, \nu_\phi), 
    \;\;\;\;
    \gamma_{cj} \sim \mbox{Beta}(a_{\gamma}, b_{\gamma}),
    \;\;\;\;
    \tau_c \sim \mbox{Beta}(1, \nu_\tau)
\end{align*}
where $\delta_{ckj}$ is a binary indicator specifying whether $p(X_{ij} | Y_i = c, Z_i = k)$ is affected by the latent class membership or takes the baseline response probability $\gamma_{cj}$. For each cause $c$, this prior encourages the associated latent classes $k = 1, ..., K$ to have similar response profiles that differ only in a subset of symptoms $\{j, \delta_{ckj} \neq 0\}$. Thus it significantly improves the efficiency of the latent representation when $K$ is large. In the original proposal of \citet{zhou2015bayesian}, the baseline response vector is set to fixed values in advance. Here we model the baseline response probabilities $\gamma_{cj}$ in a data adaptive fashion, as the baseline distribution of $p(X_{ij} | Y_i = c)$ is generally unknown and hard to specify. In certain situations, we may have external information on such conditional distributions in the form of physician provided domain knowledge \citep{mccormick2016probabilistic,li2020using} which may be used to construct informative priors on $\gamma_{cj}$.  Here we use non-informative priors with $\nu_\phi = a_\gamma = b_\gamma = 1$. 

We also expect deaths due to different causes to exhibit varying levels of complexity in the reported symptoms. For example, distribution of symptoms among deaths from external causes (e.g., drowning) might require fewer parameters to characterize than those from deaths due to infectious diseases. Instead of specifying different latent class size $K$ for each cause, the complexity of the latent symptom profiles are governed by the cause-specific sparsity level $\tau_c$. We use a non-informative prior on $\tau_c$ with $\nu_\tau = 1$. We note that the sparsity induced by the mixture prior on $\btheta$ does not perform variable selection, but rather it aims to reduce the number of parameters by identifying symptoms that are conditionally independent of others. When $\{\delta_{c1j}, ..., \delta_{cKj}\}$ are all $0$'s for symptom $j$ and cause $c$, it implies that symptom $j$ is independent of all other symptoms given cause $c$. The conditional independence model assumed by the InSilicoVA algorithm can be thought of as the special case where $\bdelta = \bm 0$.




To complete the hierarchical specification with priors for the latent parameters, we consider the stick-breaking prior for the mixing weights $\blambda$ and a Dirichlet prior on $\bpi$, i.e., 
\begin{align*}
&\lambda_{ck} = V_{ck} \prod_{l < k} (1 - V_{cl}),  \;\;
V_{ck} \sim \mbox{Beta}(1, \omega_c), \\
&\omega_c \sim \mbox{Gamma}(a_\omega, b_\omega),  \;\;\; 
\bpi \sim \mbox{Dirichlet}(\balpha).
\end{align*}

With a training dataset $\TT = \{(\bx_1, y_1), ..., (\bx_n, y_n)\}$, the posterior predictive distribution for the cause of death of a single new data point $(X^{(0)}, Y^{(0)})$ from the same population can be readily computed by 
\[
p(Y^{(0)} = c | X^{(0)}, \TT) = \int p(Y^{(0)} = c | X^{(0)}, \bpi, \blambda, \btheta) p(\bpi, \blambda, \btheta | \TT) d(\bpi, \blambda, \btheta). 
\]
VAs are most commonly used to determine the CSMF of a target population. 
Consider a target dataset $\TT_0 = \{\bx_1^{(0)}, ..., \bx_{n^{(0)}}^{(0)}\}$ with unknown cause of death, we assume the same data generating process with target-specific distributions of causes of death and latent classes. For $i = 1, ..., n^{(0)}$, we let 
\begin{align*}
    Y_i^{(0)}   &\sim \mbox{Cat}(\bpi^{(0)}), \\
    Z_i^{(0)} | Y_i^{(0)} = c  &\sim \mbox{Cat}(\blambda^{(0)}_{c}),\\
    X_{ij}^{(0)} | Y_i^{(0)} = c, Z_i^{(0)} = k &\sim \mbox{Bern}(\theta_{ckj}).
\end{align*}

For the prior on the target CSMF, we let $\bpi^{(0)} \sim \mbox{Dirichlet}(\balpha^{(0)})$. We consider two different specifications for $p_{\TT_0}(X|Y)$:
\begin{enumerate}
  \item \emph{Constant weights}: The most restrictive parameterization is to assume only \emph{prior shift} exists across the domains. That is, we let $p_{\TT_0}(X | Y) = p_{\TT}(X | Y)$, or equivalently, $\blambda^{(0)} = \blambda$.

  \item \emph{New weights}: We may relax the \emph{prior shift} assumption by allowing the target dataset to have a different composition over the latent classes learned in the training domain. In other words, we assume $p_{\TT_0}(X | Y, Z) = p_{\TT}(X | Y, Z)$ only. The new mixing weights can be estimated with the same prior as in the first stage, i.e., 
  \begin{align*}
    &\blambda^{(0)}  \sim V^{(0)}_{ck} \prod_{l < k} (1 - V^{(0)}_{cl}),  \;\;
    V^{(0)}_{ck} \sim \mbox{Beta}(1, \omega^{(0)}_c),   \;\;
    \omega^{(0)}_c \sim \mbox{Gamma}(a_\omega, b_\omega)
    \end{align*}
\end{enumerate}

\rev{We note that the additional flexibility of the second predictive strategy may sometimes suffer from overfitting in the target domain when $K$ is large. With data from multiple domains, we could further estimate $\blambda^{(0)}$ in a more robust fashion, as described in the next subsection.}

\subsection{The multi-domain model}
\label{sec:multi-model}
When training data are collected from a single domain, only one set of mixing weights can be inferred. Thus the mixing weights in the target domain are treated either as the same as those from the training domain, or as independent random variables. When the training data consist of observations from multiple domains, however, we can leverage the estimated collection of mixing weights to improve out-of-domain prediction.
Consider a training dataset consisting of observations from $G$ distinct domains. We use $D_i$ to denote the domain membership for the $i$-th observation so that $D_i \in \{0, ...,  G\}$ with $D_i = 0$ indicating the target domain. The unified data generating process for all domains can be expressed as 
\begin{align*}
    Y_i | D_i = g &\sim \mbox{Cat}(\bpi^{(g)}), \\
    Z_i | Y_i = c, D_i = g &\sim \mbox{Cat}(\blambda_c^{(g)}),\\
    X_{ij} | Y_i = c, Z_i = k &\sim \mbox{Bern}(\theta_{ckj}).
\end{align*}
That is, we assume there are $K$  latent symptom profiles nested within each of the $C$ causes across all domains, but the distribution of the causes and latent classes are domain-specific. The domain-specific distributions of $p(X|Y)$ illustrated earlier can then be explained by different mixing over latent classes. Similar to the single-domain model, we put independent stick breaking priors on the domain-specific weights and Dirichlet priors for the CSMF, i.e., for $g = 1, ..., G$,
\begin{align*}
&    
\lambda^{(g)}_{ck} = V^{(g)}_{ck} \prod_{l < k} (1 - V^{(g)}_{cl}),   \;\;
V^{(g)}_{ck} \sim \mbox{Beta}(1, \omega^{(g)}_{c}),\\
&
\omega^{(g)}_{c} \sim \mbox{Gamma}(a_\omega, b_\omega),  \;\;
\bm\pi^{(g)} \sim \mbox{Dirichlet}(\balpha^{(g)}).
\end{align*}

When performing cause-of-death assignment for a target domain without labeled death, the estimation of the mixing weights $\blambda^{(0)}$ should ideally borrow information from the collection of mixing weights $\{\blambda^{(g)}\}_{g = 1, ..., G}$ in the training domains. 
For the target domain, we let
$
\lambda^{(0)}_{ck} = \sum_{g=1}^G \eta_{cg}\lambda^{(g)}_{ck},
$
i.e., we assume the deaths due to each cause from the new domain can be modeled as a mixture of deaths due to the same cause from the observed domains. 
Compared to putting independent priors on $\blambda^{(0)}$ and estimating it without constraints, this parameterization allows the cause-of-death classification to be less sensitive to small and noisy target domain data.
We note that this modeling assumption can be relaxed if domain-level covariates are known and can be used to model domain similarities. Here we consider the more common case where no additional information about the target domain exists and leave covariate-based modeling to future work. 
Specifically, we  consider the following two strategies to parameterize $\blambda^{(0)}$, and a graphical model representation of the model is summarized in Figure \ref{fig:dag}.

\begin{enumerate}
  \item \emph{Domain-level mixture}: We fix $\bmeta_c = \bmeta$ to be the same for each cause $c = 1, ..., C$ and let 
  $
      \bmeta \sim \mbox{Dirichlet}(\balpha_\eta).
  $
  That is, the relationship between the new mixing weights $\blambda^{(0)}$ and the existing weights $\{\blambda^{(1)}, ..., \blambda^{(G)}\}$ do not vary across causes of death.

  \item \emph{Domain-cause-level mixture}: The simple domain-level mixture formulation can be too restrictive when the cause-of-death distributions are unbalanced, as the symptom distributions in the target domain are assumed to be similar to the domains with large $\eta$ conditional on any cause, including those rarely observed in these domains. Instead, we can perform cause-specific partial pooling by letting $
    \bmeta_c \sim \mbox{Dirichlet}(\balpha^{(c)}_{\eta}). 
    $
    We specify the concentration parameters to be  $\balpha^{(c)}_{\eta} = \balpha_\eta m_{cg}$, where $m_{cg}$ is the fraction of deaths from domain $g$ among all deaths due to cause $c$. This domain-cause-level mixture encourages $\blambda^{(0)}_c$ to be similar to the corresponding weights in domains with larger sample sizes for cause $c$. When domain-level covariates exist, $\balpha^{(c)}_{\eta}$ can also be parameterized as proportional to other domain-level similarity measures. 
\end{enumerate}

\rev{
\begin{figure}[!htb]
\centering
\resizebox{\textwidth}{!}{
\begin{tikzpicture}
\tikzstyle{main}=[circle, minimum size = 7mm, thick, draw =black!80, node distance = 16mm]
\tikzstyle{large}=[circle, minimum size = 12mm, thick, draw =black!80, node distance = 16mm]
\tikzstyle{connect}=[-latex, thick]
\tikzstyle{box}=[rectangle, draw=black!100]

  \node[main] (alpha_g){$\alpha^{(g)}$};
  \node[main] (pi_g) [right=1cm of alpha_g] {$\bm\pi^{(g)}$};    
  \path (alpha_g) edge [connect] (pi_g);

  \node[main] (omega_c) [below= 0.7cm of alpha_g] {$\omega_c^{(g)}$};
  \node[main] (lambda_cg) [right=1cm of omega_c] {$\bm\lambda_c^{(g)}$};
  \path (omega_c) edge [connect] (lambda_cg);

  \node[rectangle, inner sep=0mm, fit= (omega_c) (lambda_cg),label=below right:{$c = 1, ..., C$}, xshift=-6mm, yshift = 1mm] {};    
  \node[rectangle, inner sep=4mm,draw=black!100, fit= (omega_c) (lambda_cg)] (rec_omega_c){};
  \node[rectangle, inner sep=0mm, fit= (alpha_g) (rec_omega_c),label=below right:{$g = 1, ..., G$}, xshift=-14mm] {};    
  \node[rectangle, inner ysep=4.6mm, inner xsep=2.6mm,draw=black!100, fit= (alpha_g) (rec_omega_c)] {};
  \node[main] (a_omega) [above left=0.2cm and 1.5cm of omega_c] {$a_\omega$};
  \node[main] (b_omega) [below left=0.2cm and 1.5cm of omega_c] {$b_\omega$};
  \path (a_omega) edge [connect] (omega_c);
  \path (b_omega) edge [connect] (omega_c);

  \node[large] (y1) [right= 3.5cm of pi_g, fill = gray!20] {\Large $y_i$};    
  \path (pi_g) edge [connect] (y1);
  \node[large] (z1) [below left = 1cm and 0.5cm of y1] {\Large $z_i$};    
  \path (lambda_cg) edge [connect] (z1);
  \path (y1) edge [connect] (z1);
  \node[large] (x1) [right=of z1, fill = gray!20] {\Large $x_i$};    
  \path (z1) edge [connect] (x1);
  \path (y1) edge [connect] (x1);
  \node[rectangle, inner sep=0mm, fit= (z1) (y1) (x1),label=below right:{$i \in \mathcal{T}$}, xshift=-2mm] {};    
  \node[rectangle, inner sep=4.6mm,draw=black!100, fit= (z1) (y1) (x1)] (rec_omega_c){};

  \node[large] (z0) [below= 2cm of z1] {\Large $z_i$};    
  \node[large] (y0) [below right= 1cm and 0.5cm of z0] {\Large $y_i$};    
  \path (y0) edge [connect] (z0);
  \node[large] (x0) [right=of z0, fill = gray!20] {\Large $x_i$};    
  \path (z0) edge [connect] (x0);
  \path (y0) edge [connect] (x0);
  \node[rectangle, inner sep=0mm, fit= (z0) (y0) (x0),label=below right:{$i \in \mathcal{T}_0$}, xshift=-5mm] {};    
  \node[rectangle, inner sep=4.6mm,draw=black!100, fit= (z0) (y0) (x0)] (rec_omega_c){};

  \node[main] (lambda_0) [left=1.9cm of z0] {$\bm\lambda_c^{(0)}$};    
  \node[main] (eta) [left= 1cm of lambda_0] {$\bm\eta_c$};
  \node[main] (pi_0) [left=3.3cm of y0] {$\bm\pi^{(0)}$};    
  \node[main] (alpha_0) [left= 1.2cm of pi_0] {$\alpha^{(0)}$};
  \path (alpha_0) edge [connect] (pi_0);
  \path (eta) edge [connect] (lambda_0);
  \path (lambda_cg) edge [connect] (lambda_0);
  \path (lambda_0) edge [connect] (z0);
  \path (pi_0) edge [connect] (y0);
  \node[rectangle, inner sep=0mm, fit= (eta) (lambda_0),label=below right:{$c = 1, ..., C$}, xshift=-6mm] {};    
  \node[rectangle, inner sep=4.6mm,draw=black!100, fit= (eta) (lambda_0)] (rec_omega_c){};

  \node[main] (phi) [above right = 0.5cm and 2.5cm of x1]{$\phi_{ckj}$};
  \node[main] (delta) [right= 0.5cm of phi]{$\delta_{ckj}$};
  \node[main] (theta) [below=1.2cm of phi]{$\theta_{ckj}$};
  \path (delta) edge [connect] (theta);
  \path (phi) edge [connect] (theta);    
  \node[main] (tau) [right = 1cm of delta]{$\tau_c$};
  \path (tau) edge [connect] (delta);    
  \node[main] (nu_tau) [above = of tau]{$\nu_\tau$};
  \path (nu_tau) edge [connect] (tau);    
  \node[main] (nu_phi) [left = 1.5cm of nu_tau]{$\nu_\phi$};
  \path (nu_phi) edge [connect] (phi);    
  \node[main] (gamma) [below = 1.2cm of theta]{$\gamma_{cj}$};
  \path (gamma) edge [connect] (theta);
  \node[main] (agamma) [below left = 2cm and 0.2cm of gamma]{$a_{\gamma}$};
  \node[main] (bgamma) [below right = 2cm and 0.2cm of gamma]{$b_{\gamma}$};
  \path (agamma) edge [connect] (gamma);
  \path (bgamma) edge [connect] (gamma);
  \path (theta) edge [connect] (x1);
  \path (theta) edge [connect] (x0);
  \node[rectangle, inner sep=0mm, fit= (delta) (theta),label=below right:{$k = 1, ..., K$}, xshift=-17mm, yshift = -0.3cm] {};    
  \node[rectangle, inner sep=4.6mm,draw=black!100, fit= (delta) (theta)] (rec_k){};
  \node[rectangle, inner sep=0mm, fit= (delta) (gamma),label=below right:{$j = 1, ..., p$}, xshift=-12mm, yshift = -1.7cm] {};    
  \node[rectangle, inner sep=7.5mm,draw=black!100, fit= (delta) (gamma)] (rec_j){};
  \node[rectangle, inner sep=0mm, fit= (tau) (rec_j),label=below right:{$c = 1, ..., C$}, xshift=-18mm, yshift = -1cm] {};    
  \node[rectangle, inner xsep=2.5mm, inner ysep=3.5mm, yshift = -2mm, draw=black!100, fit= (tau) (rec_j)] (rec_c){};

\end{tikzpicture}
}
\caption{Graphical model of our Bayesian nested latent model. Boxes indicate replication of random variables and shaded nodes the observations. The middle block correspond to observed and latent variables for each death. The right block correspond to the shared parameters describing the symptom profile collections. The left block correspond to parameters describing the site-level mixing of the latent symptom profiles.  }
\label{fig:dag}
\end{figure}
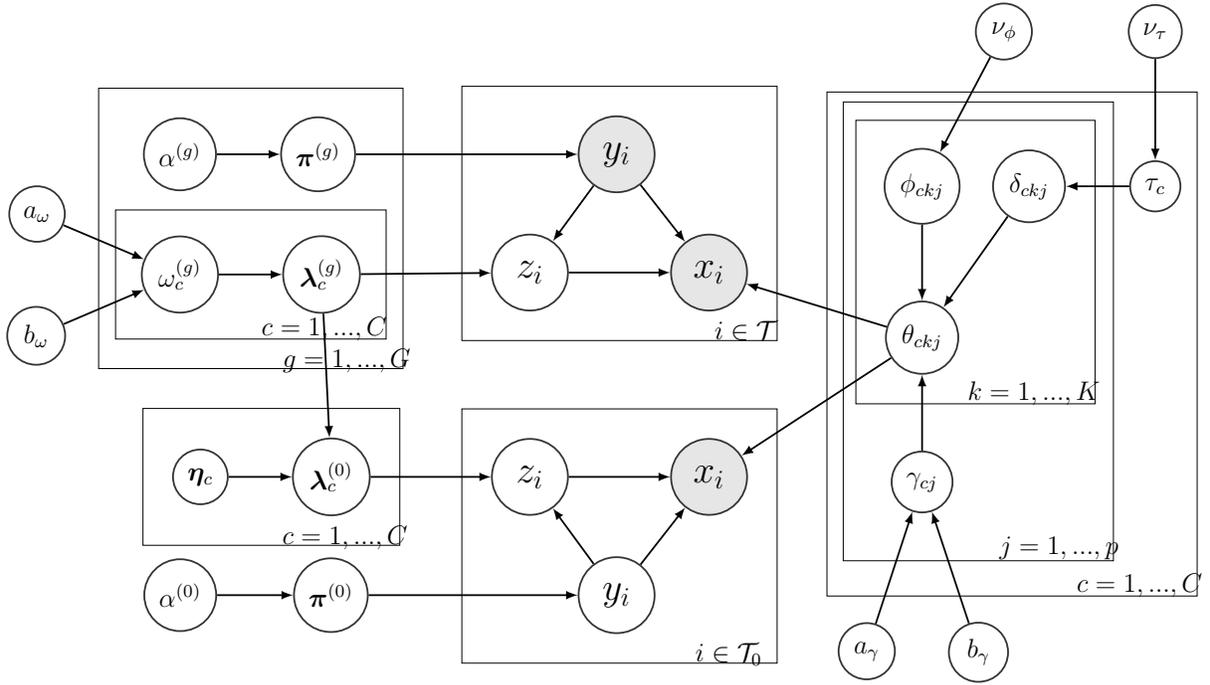
}

With a sufficient number of latent classes, both the single-domain and multi-domain models can capture the data distribution in the training data equally well. The advantage of the multi-domain model, however, is in the out-of-domain prediction, as $\blambda^{(0)}$ can be estimated with more flexibility. The similarity parameter, $\bmeta$, quantifies the similarity between the target and existing training domains in terms of the decomposition of latent classes within causes, as illustrated in Section \ref{sec:results-phmrc}.

\subsection{Practical considerations}
\label{sec:modular}
In the context of VA, $\btheta$ serves as the role of symptom-cause-information (SCI) \citep{Clark2018} on the latent space and is assumed to be shared across domains. In our implementation, we fix the distribution of $\btheta$ to its posterior distribution given the training data $\TT$ when modeling the VAs collected in the target domain. This modular procedure has several advantages. First, it prevents the information on $\btheta$ from flowing from the target domain to the training domains, as the target domain typically have few or no labeled deaths. This type of modularized procedure is known as `cutting the feedback' \citep{plummer2015cuts} and has been studied extensively in different applications \citep[e.g.]{lunn2009combining,cucala2009bayesian,zigler2013model,jacob2017better,pompe2021asymptotics}. In addition, a naive joint model likely will lead to the unlabeled deaths forming their own mixtures away from the latent classes in the training domains \citep{adams2009archipelago}. Second, as the information from the training domains can be summarized in the posterior distributions of $(\btheta, \blambda)$, users of the algorithm do not need to have access to the entire training dataset, which could be subject to data sharing restrictions. Third, the modular approach allows us to improve the exploration of parameter space by running multiple chains in parallel in the training stage, leading to a better approximation of $p(\btheta | \TT)$ and reduced computational cost for practitioners.

Lastly, data collected through VA inevitably contain a high proportion of missing responses due to the respondents not knowing whether certain symptoms exist or not being willing to talk to the interviewer about them. Some of the missing symptoms can be deterministically inferred by the logic of the questionnaire whereas most of them are difficult to impute systematically. We assume these missing responses are missing at random, i.e., the probability of the missing data mechanism depends on observed data but not on missing values. This is a standard assumption made by existing VA algorithms \citep{mccormick2016probabilistic,tsuyoshi2017,moran2021bayesian} and allows us to avoid building another imputation model, which is extremely challenging to specify and validate in practice. Under the missing at random assumption, we can conduct inference on model parameters using only the observed data.
More discussions on this assumption can be found in \citet{tsuyoshi2017}.



\subsection{Related work outside of VA}
\rev{
Learning from data collected in different domains is an active area of research in machine learning and has been explored in various applications including natural language processing \citep{ramponi2020neural}, visual classification \citep{wang2018deep}, sentiment prediction \citep{glorot2011domain}, and more recently in prediction problems in public health and clinical settings \citep{rehman2018domain,mhasawade2020population,laparra2020rethinking}. The task we consider in this paper is usually categorized as unsupervised domain adaptation (UDA), where no labeled data exist in the target domain. In particular, our estimation framework can be seen as a domain generalization method \citep{muandet2013domain} as the training stage does not involve the target domain data. A comprehensive review of UDA can be found in \citet{wilson2020survey}. Most of the UDA methods rely on creating domain-invariant representations of the features \citep[e.g.,][]{tzeng2014deep,shen2022connect}. There is extensive literature focusing on using deep learning to extract such representation in computer vision and natural language processing, but these methods are less appropriate in the context of VA as they require extensive training data \citep{oquab2014learning}. Instead, we take a flexible parametric approach and learn a parsimonious and interpretable symptom representation that is shared across domains. 

Bayesian hierarchical models are routinely used to model heterogeneous, multi-domain data in many scientific fields, though less common in the domain adaptation literature. \citet{bruzzone2001unsupervised} and \citet{raghuram2012semisupervised} proposed Gaussian mixture models to learn the data representation of training domains and apply to target domains in remote sensing. \citet{wood2009hierarchical} proposed nonparametric Bayesian methods for domain adaptation in language models. \citet{hajiramezanali2018bayesian} and \citet{boluki2021optimal} developed Bayesian hierarchical models for modeling RNA sequencing data from multiple domains. While being tailored for different applications, the common conceptual framework is to model the data distribution in different domains as mixtures of shared latent clusters, an insight that we also use in building our proposed latent class model.   
}

\section{Posterior inference}
\label{sec:sampling}

We have the following posterior joint distribution given the training dataset $\TT = \{(\bx_i, y_i, d_i)\}_{i=1, ..., n}$,
\begin{align*}
p(\bpi, \blambda, \bphi, \bgamma, \bomega | \TT)
\propto & 
\prod_{i = 1}^{n} p(y_i | \bpi^{(d_i)})\sum_{z_i=1}^K \Big( p(z_i | y_i, \blambda^{(d_i)})\prod_{j = 1}^p p(x_{ij} | y_i, z_i, \bgamma, \bphi, \bdelta) \Big)\\
& \times
\prod_{c = 1}^C\prod_{j = 1}^p  p(\gamma_{cj})
\prod_{c = 1}^C\prod_{k = 1}^K\prod_{j = 1}^p p(\phi_{ckj})p(\delta_{ckj} | \tau_c)
\prod_{c = 1}^C p(\tau_c)
\\
& \times \prod_{g = 1}^G p(\bpi^{(g)})p(\blambda^{(g)} | \bomega^{(g)})p(\bomega^{(g)}).
\end{align*}

The posterior distribution of the single-domain model is a special case with $G = 1$ and thus omitted. In the target domain, the modularized joint posterior distribution is 
\begin{align*}
  p(Y^{(0)}, \bpi^{(0)} | X^{(0)}, \TT) \propto 
&  
  \int\prod_{i = 1}^{n^{(0)}} p(y_i^{(0)} | \bpi^{(0)})\sum_{z_i^{(0)}=1}^K \Big( p(z_i^{(0)} | y_i^{(0)}, \blambda^{(0)})\prod_{j = 1}^p p(x_{ij}^{(0)} | y_i^{(0)}, z_i^{(0)}, \btheta) \Big)
\\
& 
p(\bpi^{(0)})
p(\blambda^{(0)} | \blambda^{(1)}, ..., \blambda^{(G)}, \bmeta) 
p(\bmeta)
\\
& 
p(\blambda^{(1)}, ..., \blambda^{(G)},  \btheta | \TT) d(\bmeta, \blambda^{(0)}, ..., \blambda^{(G)},  \btheta).
\end{align*}

The posterior distribution of the model parameters is not available in closed form, but we can easily draw posterior samples from a Gibbs sampler. The detailed procedure of fitting the model is described next.


\subsection{Training stage} We first infer the parameters associated with the training data using the following steps. 
\begin{enumerate}
  \item Sample the latent class membership $Z_i$ for $i \in \TT$ with 
  \[
      p(Z_{i} = k | Y_i = c, D_i = g, \btheta, \blambda) \propto \lambda^{(g)}_{ck}\prod_j \theta_{ckj}^{x_{ij}}(1-\theta_{ckj})^{1-x_{ij}}.
  \]
  \item Sample the stick-breaking parameters $\bV^{(g)}$ and $\omega^{(g)}$ for $g = 1,..., G$. Denote $n^{(g)}_{ck} = \sum_i \bm 1_{Y_i = c, Z_i = k, D_i = g}$. The full conditional is 
  \begin{align*}
  V_{ck}^{(g)} | Y, Z, \bomega^{(g)} &\sim \mbox{Beta}(1 + n_{ck}^{(g)}, \omega_c^{(g)} + \sum_{l > k} n^{(g)}_{cl}), 
  \;\; c = 1, ..., C; k = 1, ..., K-1,
  \\
  \omega_c^{(g)} | \bV^{(g)} &\sim \mbox{Gamma}(a_\omega + K - 1, b_\omega - \sum_{k=1}^{K-1} \log(1-V_{ck}^{(g)})),
  \;\; c = 1, ..., C.
  \end{align*}

  \item Sample the response probabilities $\btheta$ after integrating out the latent binary indicators $\bdelta$. The posterior conditional of $\theta_{ckj}$ is a mixture of Beta distribution and point mass at the baseline vector $\gamma_{cj}$.  We let $n_{ckj1} = \sum_i \bm 1_{Y_i = c, Z_i = k, X_{ij} = 1}$ and $n_{ckj0} = \sum_i \bm 1_{Y_i = c, Z_i = k, X_{ij} = 0}$. The posterior conditional of $\theta_{ckj}$ is
  \[
    p(\theta_{ckj} | \cdot) = \tilde \pi_{ckj}\mbox{Beta}(1 + n_{ckj1}, \nu_\phi + n_{ckj0}) + (1-\tilde \pi_{ckj})I(\gamma_{cj}), 
  \]
  where $I(\gamma_{cj})$ is the point mass at $\gamma_{cj}$. The mixing probability is 
  \[
      \tilde \pi_{ckj} = \frac{\tau_cB(1+n_{ckj1}, \nu_\phi + n_{ckj0})}{
      \tau_c B(1+n_{ckj1}, \nu_\phi + n_{ckj0})
      + 
      (1-\tau_c) B(1, \nu_\phi)
      \gamma_{cj}^{n_{ckj1}}
      (1-\gamma_{cj})^{n_{ckj0}}
      }.
  \]
  \item Sample the latent binary indicators $\bdelta$ from its full conditional,
  $$
    \delta_{ckj} \sim \mbox{Bern}(\tilde\pi_{ckj}).
  $$
  \item Sample the sparsity level parameter $\tau_c$,
  $$
  \tau_c \sim \mbox{Beta}(1 + \sum_{j}\sum_{k}\delta_{ckj}, \nu_\tau + \sum_{j}\sum_{k}(1 - \delta_{ckj})).
  $$

  \item Sample the baseline vector $\bgamma$ from its full conditionals. Let $O_j = \{i, x_{ij} \textnormal{ not missing}\}_{i = 1, ..., n}$, we have
  \[
      \gamma_{cj} | X, Y, Z, \bdelta \sim \mbox{Beta}(a_\gamma + \sum_{i\in O_j} x_{ij}(1-\delta_{y_iz_ij}),  b_\gamma + \sum_{i\in O_j} (1 - x_{ij})(1-\delta_{y_iz_ij})), 
  \]

    \item {Update the CSMF vector in each domain $\bpi^{(g)}|Y$.} Let $n^{(g)}_c = \sum_{i} \bm 1_{Y_i = c, D_i = g}$,
  \[
     \bpi^{(g)} | Y \sim \mbox{Dirichlet}(\alpha_{\pi, 1} + n^{(g)}_1, ..., \alpha_{\pi, C} + n^{(g)}_C).
  \]

\end{enumerate}

When there exist observations from the training domains with unknown labels, the above posterior sampling procedure can be easily modified by updating the cause-of-death assignment and latent class indicator for each unlabeled data with 
\[
  p(Y_i  = c, Z_i = k| X_i,  D_i = g) \propto   \pi_c^{(g)} \lambda_{ck}^{(g)}\prod_j (\theta_{ckj})^{x_{ij}}(1-\theta_{ckj})^{1-x_{ij}}.
\]

\subsection{Prediction stage}
When performing cause-of-death assignment for the target dataset, we take the modular approach described in Section \ref{sec:modular} by plugging in the posterior draws of $\btheta$ and $\blambda$ from the training stage model. We describe the sampler for the other parameters in the multi-domain model below and leave the case of single-domain model to the Supplementary Materials.

\begin{enumerate}
\item Sample the cause-of-death assignments $Y_i$ and latent class indicator $Z_i$. To facilitate easier computation, we augment the data with latent indicator $\tilde D_i \in \{1, ..., G\}$ for observations from the target population and sample all latent indicators jointly with 
\[
  p(Y_i  = c, Z_i = k, \tilde D_i = g | X_i, D_i = 0, \blambda, \btheta) 
  \propto 
  \pi_c^{(0)}\eta_{cg} \lambda_{ck}^{(g)}\prod_j (\theta_{ckj})^{x_{ij}}(1-\theta_{ckj})^{1-x_{ij}}.
\]
 \item For the multi-domain model with domain-level mixture, update the domain similarity weights $\bmeta$ with
  \[
  \bmeta | Y, \tilde D \sim \mbox{Dirichlet}(\alpha_{\eta, 1} + \sum_{i} \bOne_{\tilde D_i = 1}, ..., \alpha_{\eta, G} + \sum_{i} \bOne_{\tilde D_i = G}).
  \]
 For the multi-domain model with domain-cause-level mixture, 
  \[
    \bmeta_c | Y, \tilde D \sim \mbox{Dirichlet}(\alpha_{\eta, 1}m_{c1} + \sum_{i} \bOne_{Y_i = c, \tilde D_i = 1}, ..., \alpha_{\eta, G}m_{cG} + \sum_{i} \bOne_{Y_i = c, \tilde D_i = G}).
  \]

  \item Update the CSMF vector in the target domain $\bpi^{(0)}|Y$ with
  \[
     \bpi^{(0)} |Y \sim \mbox{Dirichlet}(\alpha_{\pi, 1}^{(0)} + n^{(0)}_1, ..., \alpha_{\pi, C}^{(0)} + n^{(0)}_C)
  \]
  where $n^{(0)}_c = \sum_{i} \bm 1_{Y_i = c, D_i = 0}$.
 
\end{enumerate}

\rev{
\subsection{Computational consideration}
The latent class model formulation leads to a multimodal posterior distribution that could be difficult to explore using MCMC in practice. One of the reasons is likelihood invariance to permutation of class labels. When we model the mixing weights in the target domain as a weighted average of the weights in the training domains, however, the model parameters for the target site do not depend on the labeling of latent classes, making the computation in the prediction stage invariant to the latent class labels. This allows us to implement the training stage sampler multiple times in parallel with different starting values and combine the posterior draws of $(\btheta, \blambda)$ afterwards. To further improve the predictive performance, we adopt the stacking approach proposed by \citet{yao2022stacking} to combine the multiple MCMC chains. 
}

Finally, We note that the sampler proposed in this section enjoys significant advantage over models that use latent Gaussian representations to characterize symptom dependence \citep{li2020using, tsuyoshi2017, moran2021bayesian}. The classification stage of \nameholder{} essentially reduces to allocating observations to $CK$ or $CKG$ latent classes with conditional independent response probabilities, which has the same order of computational complexity as the InSilicoVA algorithm \citep{mccormick2016probabilistic}.

\section{Results}
\label{sec:results-phmrc}
In this section, we present results of our model evaluated on the PHMRC gold-standard VA dataset \citep{murray2011population}. The PHMRC dataset has been used extensively in validating and comparing VA cause-of-death assignment methods \citep{mccormick2016probabilistic,tsuyoshi2017,moran2021bayesian,Clark2018}. It consists of $7,841$ adult deaths collected from six study sites (Andhra Pradesh, India; Bohol, Philippines; Dar es Salaam, Tanzania; Mexico City, Mexico; Pemba Island, Tanzania; and Uttar Pradesh, India). All deaths occurred in health facilities and gold-standard causes are determined based on laboratory, pathology and medical imaging findings. The cause of death is coded into $34$ categories, and we pre-process the raw dataset into $168$ binary symptoms as described in \citet{mccormick2016probabilistic}. \rev{The data processing steps are implemented using the \texttt{openVA} package \citep{li2022openva}.} 

\rev{
We present three sets of experiments in this section. We first evaluate the out-of-domain predictive performance of different models for a leave-one-site-out experiment. To further explore how these methods behave under more extreme data shift, we then construct resampled datasets based on each target domain, and evaluate the models on these replicated datasets. Finally, we consider the case where some local labeled data exist in the target domain and compare different approaches to incorporate the labeled data. In addition to the evaluation of predictive performance, we also compare computational time of different methods and discuss the estimated latent symptom profiles. We also perform comparisons on synthetic data and results of these simulations are included in the Supplementary Materials.
}

\subsection{Prediction for a new study site}
\label{sec:results-phmrc-1}
For our first evaluation, we consider the task of out-of-domain prediction for each of the six study sites. We iteratively take one site as the target domain and the rest of the five sites as training domains. A summary of the true CSMFs for each site is included in the Supplementary Materials. 

\rev{Throughout all the experiments in this paper, we use the same six sets of training data, each including five sites. Since our goal is to develop models that are robust to unseen target domains, we should choose $K$ based solely on the training data. To determine the value of $K$, we first fit the multi-domain model with a large $K$ and truncate the latent class rarely occupied. For each experiment, we first let $K = 15$ and monitor the number of latent classes occupied in the posterior samples of $Z_i$. Figure \ref{fig:chooseK} plots the fraction of posterior samples where each latent class is occupied for each cause of death in each experiment. The latent classes after the largest $10$ classes are occupied in less than $5\%$ of posterior samples across almost all causes and experiments. Therefore, we let $K = 10$ in all our subsequent analysis of the PHMRC data. The performance of \nameholder{} does not vary much from different choices of $K$. Additional sensitivity analysis are provided in the Supplementary Materials.}

\rev{
  \begin{figure}[!htb]
  \centering
  \includegraphics[width = .8\textwidth]{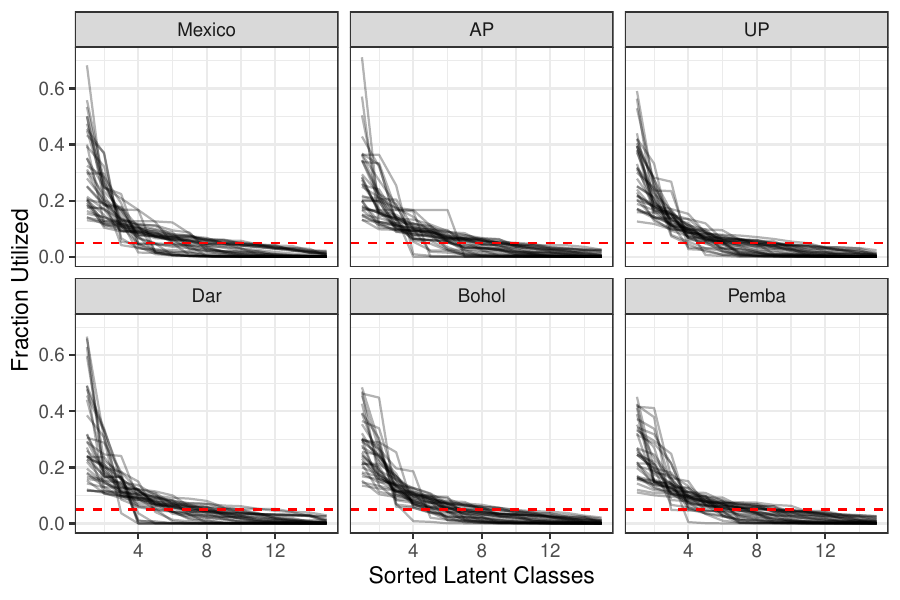}
  \caption{The fraction of latent classes utilized for each experiment. The latent classes are sorted by the fraction of posterior draws where they are occupied. Each line represents one cause of death and each panel represents a training dataset where the target site in the label is excluded.}
  \label{fig:chooseK}
  \end{figure}
}

\rev{
  We consider both the multi-domain (\nameholder{}-M) models in Section \ref{sec:multi-model}, and the single-domain (\nameholder{}-S) models in Section \ref{sec:single-model} by ignoring the domain indicators in the training data. For the concentration parameters, we let $\balpha_\pi = \balpha_\pi^{(0)} = \balpha_\eta = \bOne$. In the training stage, we ran six parallel MCMC chains with different starting values for $4,000$ iterations and discarded the first $1,000$ posterior draws as burn-in. For \nameholder{}-S with new weights, the posterior draws from a single chain were used in the prediction stage. For the other models, we re-weighted the six chains using stacking \citep{yao2022stacking} and generated $4,000$ posterior samples from the combined posterior distribution. Finally, the first half of the chain in the prediction stage were discarded as burn-in. 
}

We evaluated the performance of the proposed model based on the classification accuracy of the predicted top cause of death and the so-called `CSMF accuracy' \citep{murray2011robust}, a widely used metric to compare the estimated CSMF vector with the truth. The CSMF accuracy metric is a form of the normalized absolute error \citep{gonzalez2017review}. It measures the $L_1$ distance between the estimated and true CSMF, and is normalized to range between $0$ (worst) and $1$ (best). It is defined as
$
\mbox{CSMF}_{acc}(\hat\pi) = 1 - \frac{\sum_{c=1}^C |\hat\pi_c - \pi_c|}{2(1 - \min_{c} \pi_c)},
$ 
where $\pi_c$ is the true CSMF calculated by the empirical distribution in the test data. \rev{Unlike the top cause accuracy that focuses on individual level performance, CSMF accuracy evaluates the task of prevalence estimation, which is more informative for VA practitioners, especially in unbalanced datasets.}

We benchmarked the performance of the proposed model to the InSilicoVA algorithm. InSilicoVA is one of the most widely adopted VA methods used in practice and has been shown to have better or comparable performance compared to other VA models in use \citep{mccormick2016probabilistic,Clark2018}. We also compared with the Bayesian factor model \citep{tsuyoshi2017} and the FARVA model \citep{moran2021bayesian}. Both models approximate the joint distribution of symptoms using a latent Gaussian model and have been shown to significantly improve the predictive performance compared to InSilicoVA. For the FARVA model, we included a binary covariate indicating age at death exceeding $65$. The inclusion of covariates did not change the performance metric of FARVA by much in our experiments. We follow the recommendation by the authors of these methods in setting the tuning parameters and length of MCMC iterations. For the InSilicoVA algorithm, we ran the MCMC for $4,000$ iterations. For the Bayesian factor model, we used $6$ latent factors and obtained $1,000$ posterior samples after discarding the first $500$ samples as burn-in and saving every fifth sample. Far FARVA, we used $10$ latent factors and obtained $500$ posterior samples after discarding the first $2,000$ samples as burn-in and saving every tenth sample.


\rev{
Figure \ref{fig:phmrc-original} summarizes the top cause accuracy and CSMF accuracy for the different models evaluated on the six sites respectively. All models show significantly improved top cause accuracy compared to InSilicoVA. The CSMF accuracy of the three classes of models are also comparable and improves from the InSilicoVA results except in two cases (FARVA in Bohol and \nameholder{} in Andhra Pradesh). There is no model that dominates either predictive metrics in all six experiments. 
}

\rev{
We note, however, that an important caveat in this experiment is that the CSMFs in most sites are similar in the PHMRC dataset, except for Pemba. A comparison of the empirical CSMFs are included in the Supplementary Materials. Both the Bayesian factor model and FARVA assume a common data generating process for causes of deaths in both training and target datasets with a shared parameter $\bpi$, and thus enjoy stronger regularization from the prior compared to \nameholder{}. However, such model formulation can negatively affect the predictive performance in domains with a very different cause-of-death distribution compared to the training domains, such as Pemba in this experiment. We further evaluate the performance of the cause-of-death assignment models under more extreme scenarios in the next subsection.
}


\rev{
  \begin{figure}[htb]
  \includegraphics[width = \textwidth]{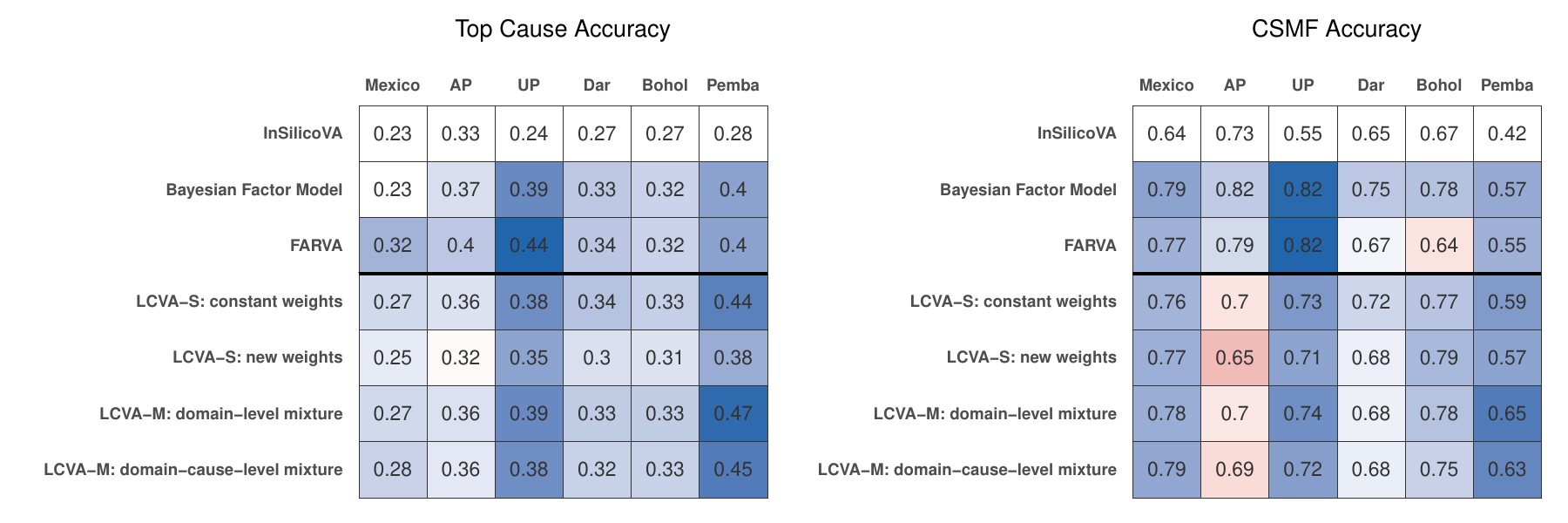}
  \caption{Individual-level top cause prediction accuracy (left) and CSMF accuracy (right) of different models. The cells are colored by the relative difference from the InSilicoVA results, where blue cells indicate higher accuracy and red cells indicate lower accuracy. All variations of \nameholder{} demonstrate good performance that is significantly higher than InSilicoVA except in one experiment. Bayesian factor model and FARVA also show comparable improvements.}
  \label{fig:phmrc-original}
  \end{figure}
}

\subsection{Prediction for a new study site under more extreme data shift}
\label{sec:results-phmrc-2}

\rev{In order to evaluate the models under more extreme data shift, for each leave-one-site-out experiment, we created $50$ synthetic target dataset by first generating $\pi^{(0)} \sim \mbox{Dirichlet}(0.1, ..., 0.1)$ and resampled deaths in the held-out site with replacement to match the generated target CSMF. This data generating process leads to the target CSMFs being concentrated on a small number of causes and thus very different from the overall prevalence in the training domains. This reflects more closely the cause-of-death distributions in small- and moderate-scale VA studies in practice. We again compare our methods to InSilicoVA, Bayesian factor model, and FARVA. The top cause accuracy and CSMF accuracy metrics vary significantly across replications due to the high variability of the target domain CSMF. Thus we focus on the relative performances of the models. 
Figure \ref{fig:phmrc-resample-1A} and \ref{fig:phmrc-resample-1B} shows the pairwise comparison of all models in terms of the fraction of times one model outperform the other in the two metrics across the six collection of target datasets. The simplest model, InSilicoVA, led to worse top cause accuracy estimation in almost all pairwise comparisons. However, in terms of the CSMF accuracy, which is more of interest for domains with unbalanced cause-of-death distribution, both Bayesian factor model and FARVA performed worse than InSilicoVA on more than half of the resampled datasets in four out of six sites. 
\nameholder{}-M with domain-cause-level mixture is the only method that outperformed InSilicoVA in both metrics on more than half of the resampled datasets across all six sites, and it outperforms all three existing methods in CSMF accuracy on more than half of the resampled datasets. \nameholder{}-M with domain-level mixture also perform very similarly. This demonstrates the robustness of the proposed models to extreme data shift. 
}
\rev{
  \begin{figure}[htb]
  \includegraphics[width = \textwidth]{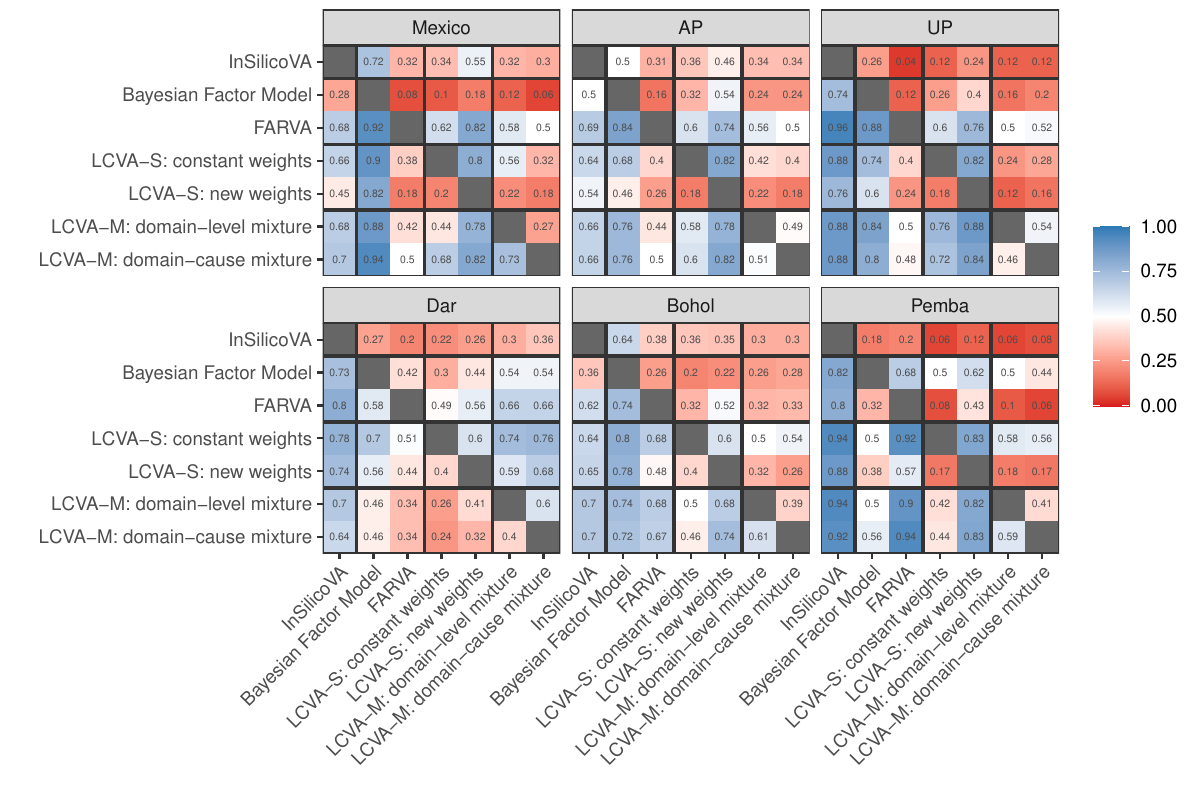}
  \caption{Pairwise comparison of all models in terms of top cause accuracy. The color of each cell represents the proportion of target datasets where the model on the left achieves higher top cause accuracy than the model on the bottom.}
  \label{fig:phmrc-resample-1A}
  \end{figure}
}

\rev{
  \begin{figure}[htb]
  \includegraphics[width = \textwidth]{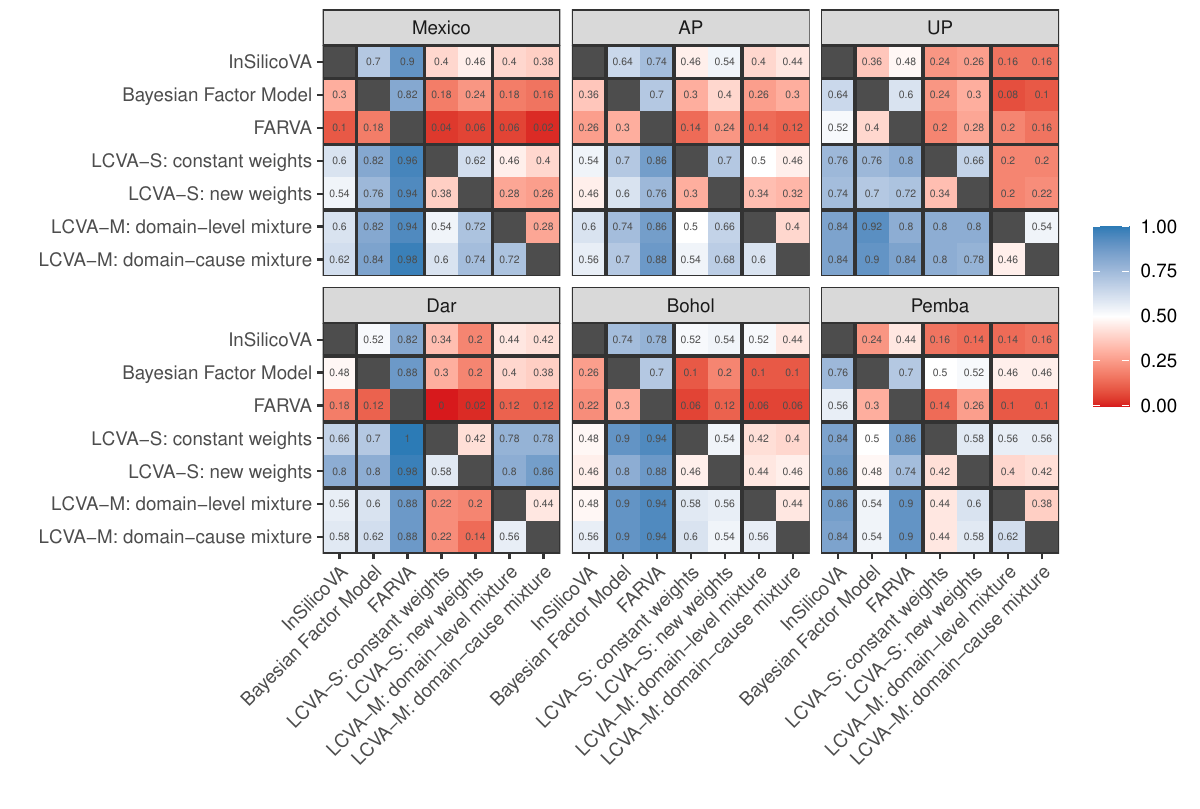}
  \caption{Pairwise comparison of all models in terms of CSMF accuracy. The color of each cell represents the proportion of target datasets where the model on the left achieves higher CSMF accuracy than the model on the bottom. }
  \label{fig:phmrc-resample-1B}
  \end{figure}
}

\rev{
We further take a closer examination of the scale of the relative improvements of the proposed models. Figure \ref{fig:phmrc-resample-2} shows the percentage improvement of each model $m$ compared to the Bayesian factor model, i.e., $(\mbox{Acc}_m - \mbox{Acc}_{BF}) / \mbox{Acc}_{BF}$ for both metrics.  In Dar es Salaam and Pemba island, \nameholder{}-M had similar performances compared to the Bayesian factor model. In the other four sites, \nameholder{}-M models achieved around $10\%$ to  $40\%$ in median improvements for top cause accuracy and $10\%$ to $20\%$ for CSMF accuracy.
FARVA also outperformed the Bayesian factor model in five sites in terms of individual-level prediction, but showed worse CSMF accuracy across all sites. Additional comparisons are included in the Supplementary Materials. 

Based on these two set of experiments and the synthetic data evaluations in the Supplementary Materials, \nameholder{}-M with domain-level mixture was shown to be more robust to different data shift scenarios while enjoying the more parsimonious parameterization than the domain-cause-level mixture model. Thus we focus our attention on this model in the rest of the paper.  
}
\rev{
  \begin{figure}[htb]
  \includegraphics[width = \textwidth]{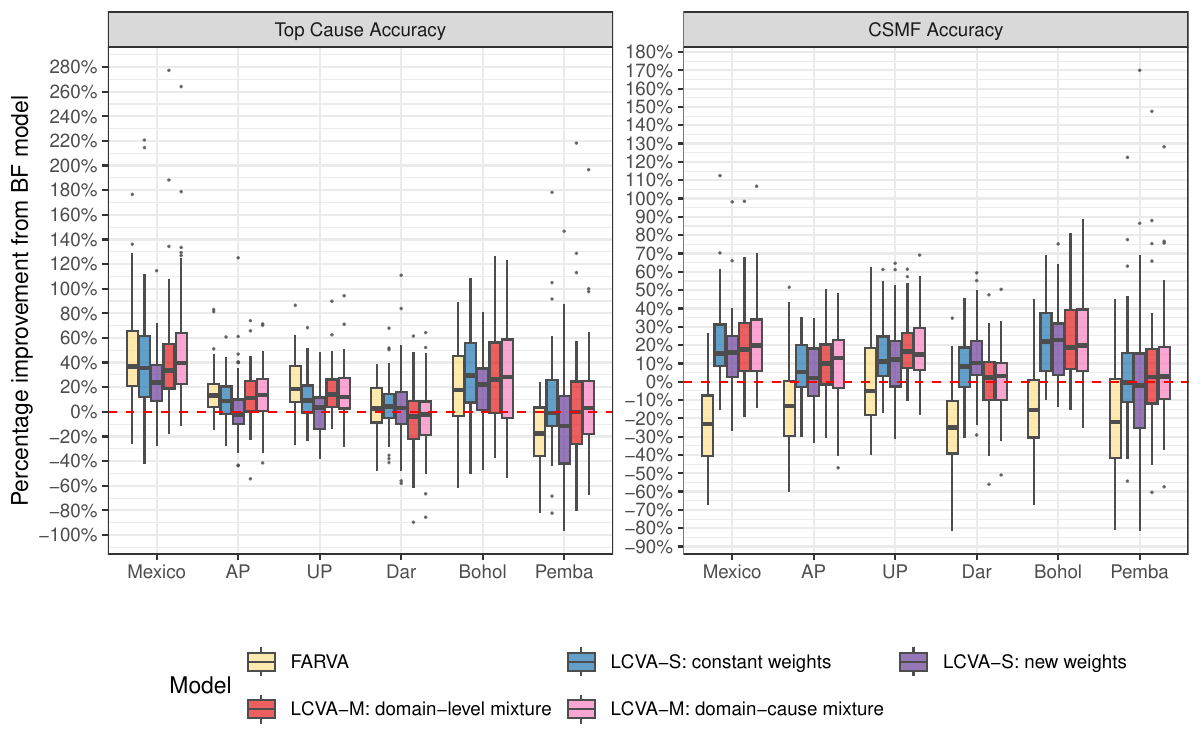}
  \caption{Boxplot of improvement over the Bayesian factor model \citep{tsuyoshi2017} in top cause accuracy (left) and CSMF accuracy (right) by different models on $50$ resampled target domain datasets based on deaths from each of the six sites.}
  \label{fig:phmrc-resample-2}
  \end{figure}
}

\subsection{Computation time}
\label{sec:results-time}

The proposed model is highly efficient in computation compared to the Bayesian factor model and FARVA. Across the six leave-one-site-out prediction experiments, 
the InSilicoVA algorithm is the fastest to implement as the symptoms are treated as conditionally independent. It required $20$ seconds to run $1,000$ iterations of MCMC.
\rev{
For the \nameholder{} with domain-level mixture, every $1,000$ iterations of MCMC required $142$ seconds in the training stage, and $42$ seconds in the prediction stage. In comparison, for the same tasks and $1000$ iterations of the MCMC, 
the Bayesian factor model used $4,275$ seconds and the FARVA model used $17,422$ seconds. Data analysis using these two methods takes hours or days to finish, which is clearly infeasible as routine tools in resource-constrained settings. The \nameholder{} requires the least additional computational cost among the methods that take into account symptom dependence.
}
The computation time are all evaluated on a MacBook Pro with 2.6 GHz 6-Core Intel Core i7 processor and 32 GB memory.

\subsection{Interpreting the latent symptom profiles}
\begin{figure}[!htp]
 \centering
\includegraphics[width = \textwidth]{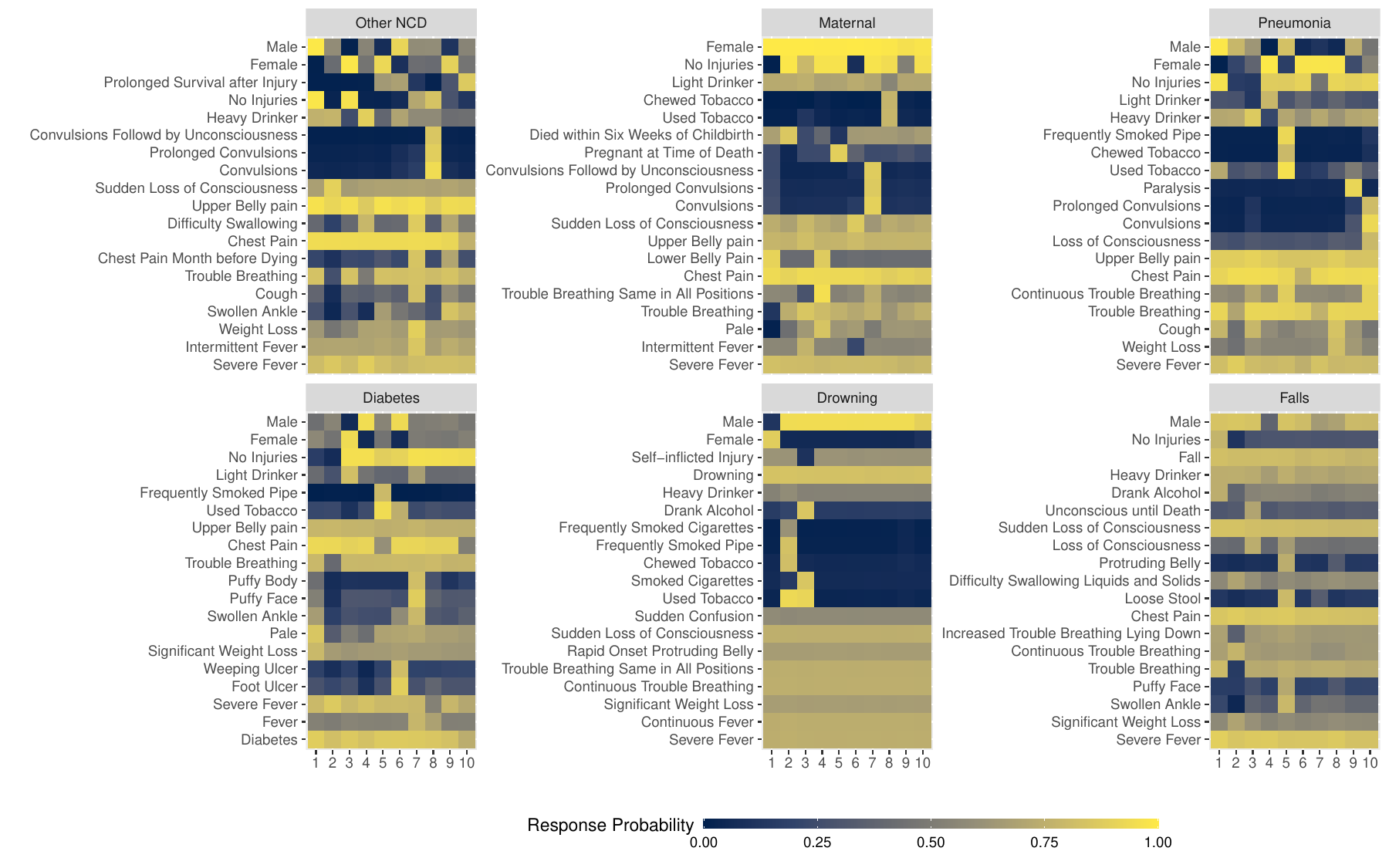}
{}
\vspace{.2cm}
\includegraphics[width = .49\textwidth]{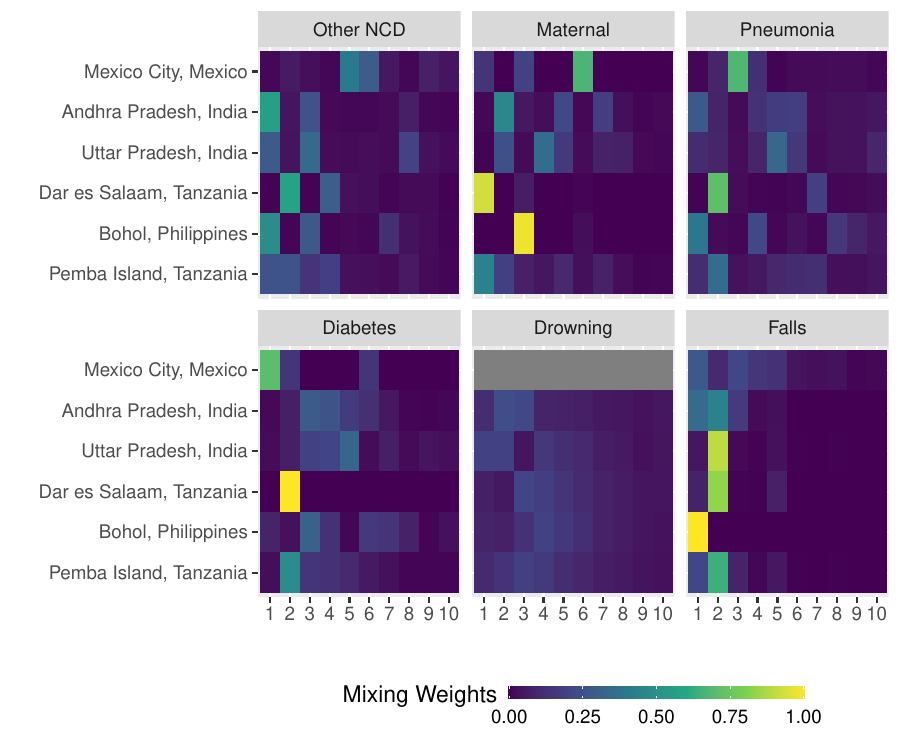}
\includegraphics[width = .49\textwidth]{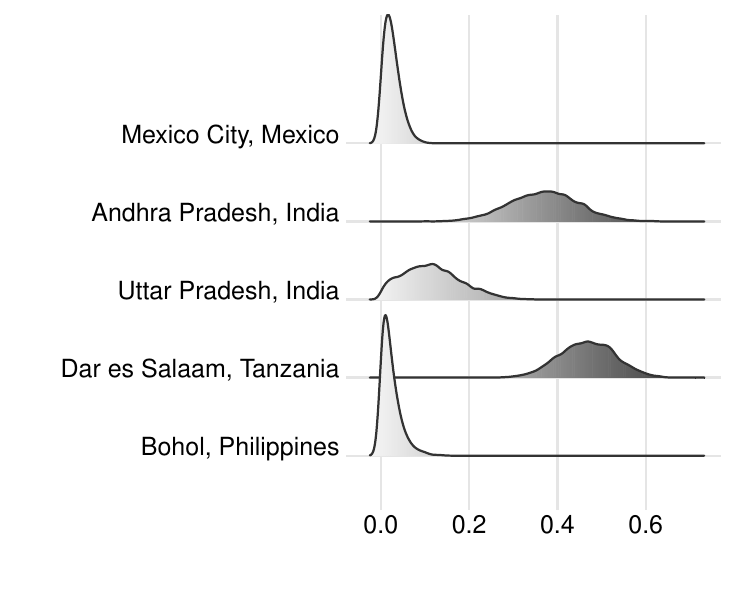}
\caption{Top: Posterior means of the conditional probabilities of observing the selected subset of symptoms given the top six causes of death in Pemba and latent class membership. Bottom: Posterior means of the mixing weights of latent classes by domain (left) and the posterior distribution of the similarity parameter $\bmeta$. Pemba island is the target domain. The mixing weights corresponding to drowning in Mexico city are removed since there is no drowning deaths in Mexico city in this dataset.}
\label{fig:pemba-response}
\end{figure}

We now turn to the estimated latent classes and domain similarity parameters. As an example, we examine the experiment treating Pemba island as the target domain and the other five larger sites as training domains. As an illustration, we focus on the multi-domain \nameholder{} with domain-level mixture model. It is worth noting that the estimated latent classes are subject to permutation and further research is needed to relate latent classes with observable characteristics of the causes of death. Nevertheless, they still reveal interesting latent structures in these data. 
The top panel of Figure \ref{fig:pemba-response} shows the estimated conditional probabilities, $\btheta$. We sort the symptoms by their largest conditional probabilities given a latent class, i.e., $\max_{k}\theta_{ckj}$, and show the conditional probabilities of the top $20$ symptoms for the six most prevalent causes in Pemba. 
\rev{Within each cause of death, most of the symptoms have similar conditional probabilities across latent classes, as encouraged by the prior on $\theta$.
Taking deaths due to `other Non-Communicable Diseases (NCD)' as an example,
the latent symptom profiles differ most significantly on the response probabilities for the sex of the respondent, injury, and specific symptoms such as convulsion, chest pain, having trouble breathing, and cough. The corresponding mixing weights are shown in the lower left panel of Figure \ref{fig:pemba-response}. 
Mexico city and Dar es Salaam have more different latent class decomposition than the other sites. For example, comparing deaths due to other NCD in Dar es Salaam with the other sites, the $1$st and $3$rd latent classes were utilized less often compared to the $2$nd and $4$th, indicating that symptoms such as having trouble breathing are less likely to be present for deaths due to other NCDs compared to other sites. 
We note, however, that such differences could be due to either different decomposition of specific NCDs or differential reporting across sites, and more studies are needed to understand their underlying mechanism. 
We also observe that for simpler causes such as drowning and falling, the latent symptom profiles are heavily shrunk towards the baseline values, indicating that the conditional independence simplification being a useful approximation for these causes. The shrinkage protects the model from overfitting on the noise in the training domains when $K$ is large for a relatively simpler causes.} Finally, the estimated $\bmeta$ in the lower right panel of Figure \ref{fig:pemba-response} shows how $\blambda^{(0)}$ is related to the existing mixing weights through the similarity parameter $\bmeta$. The domain with the most similar mixing weights to Pemba is estimated to be Dar es Salaam, which is as expected given both sites are in Tanzania.

\subsection{Incorporating labeled data from the target domain}
\label{sec:results-calib}

Finally, we consider the situation where a small subset of labeled data is available in the target domain. 
In this case, the target population  $\TT_0 = \LL \cup \UU$, where $\LL$ are the data with labels and $\UU$ are the data without labels. 
The \nameholder{} model can incorporate such information naturally by modeling all deaths in $\TT_0$ and skipping the sampling of $Y_i$ for the deaths in $\LL$. \rev{This leads to a joint modeling approach in the prediction stage in a semi-supervised fashion.} An alternative strategy is to calibrate the estimated CSMF to the known labels after fitting a model on only the training data 
and use $\LL$ to estimate the misclassification matrix and de-bias the estimated $p_{\TT_0}(Y)$ \citep{fiksel2020generalized}.

Here we again consider the six experiments of out-of-domain prediction in Section \ref{sec:results-phmrc-1}. For the calibration step, we took the posterior means of the estimated individual cause-of-death probabilities from InSilicoVA and \nameholder{}-M with domain-level mixture as the input to the calibration procedure in \citet{fiksel2020generalized}, and randomly selected $30\%$ of deaths in the target domain to reveal the true labels. 
\rev{
For each randomly selected set of labeled deaths, we compared the three estimates of the CSMFs: calibrated InSilicoVA, calibrated \nameholder{}, and \nameholder{} with partial labels treated as known in the prediction stage. The results over $50$ randomly sampled $\LL$ are summarized in Figure \ref{fig:phmrc-calibration}. 
With the additional information of the known labels, the CSMF accuracy improves for both InSilicoVA and \nameholder{}. By directly incorporating the deaths with known causes in the joint modeling approach, we were able to obtain the highest CSMF accuracy in all six sites using \nameholder{}.

We note, however, that the calibration approach in \citet{fiksel2020generalized} were developed for a much smaller cause list. The calibration procedure is less effective in this analysis due to the need of estimating a $34 \times 34$ dimensional error transition matrix using a small number of labeled data. In applications where only broader cause categories are of interest, the calibration procedure can sometimes perform similarly well compared to the joint modeling approach. We experimented with calibration using aggregated $11$ and $5$ broader causes in the Supplementary Materials. 
} 


\rev{
 \begin{figure}[!ht]
\includegraphics[width = \textwidth]{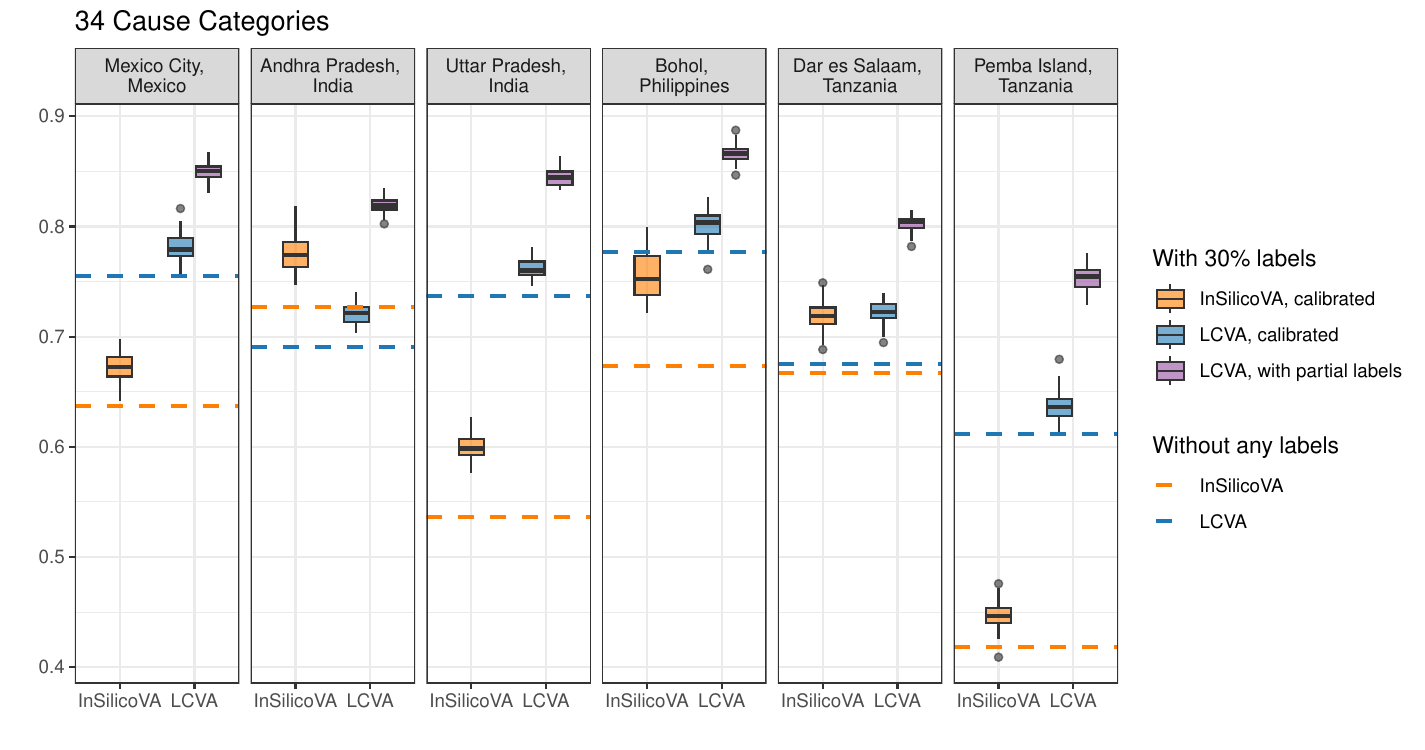}
\caption{Comparison of \nameholder{}-M with domain-level mixture and InSilicoVA when the cause of death of $30\%$ of data in the target domain are known. The horizontal line indicates the CSMF accuracy of the two methods before calibrating to the labeled deaths. The box plot shows the distribution of the CSMF accuracy over $50$ resampled labeled dataset.}
\label{fig:phmrc-calibration}
\end{figure}  
}




\section{Discussion}
\label{sec:discussion}
Quantifying the population distribution of causes of death and assigning causes to specific deaths using VAs are challenging, especially when using information derived from non-local data sources. In this paper, we proposed a statistical framework for modeling VA data collected over multiple domains and quantifying the cause-of-death distributions for deaths in a new domain. 
The proposed model framework uses a parsimonious nested latent class model to characterize the joint distributions of symptoms given causes, which allows direct interpretation of the latent structure and faster computation relative to existing methods.
We demonstrated how this framework can efficiently borrow information from multiple domains and improve out-of-domain predictive performance significantly with both synthetic and real data, \rev{and recommend a parsimonious prediction framework where we treat the mixing weights of the target domain to be a weighted average of the those in the training domains (i.e., \nameholder{}-M with domain-level mixture)}. 

The proposed method could be extended in a few different ways. 
First, 
we have treated the domains as independent datasets in this paper. In many situations, data are collected over different sub-national regions and over time. Structural information about spatial and temporal correlation, as well as domain-level similarities could further improve the modeling of domain-specific distributions, as cause of death and latent class distributions are likely to vary  smoothly across similar domains.  \rev{Summarizing domain heterogeneity and similarities using the latent class model can also be of scientific interest to understand the underlying population.} In addition, symptom distributions conditional on similar causes of death may also show similarities. Tree-based methods for domain adaptation can be a fruitful direction to explore such additional information \citep{wu2021double}.
Second, 
it is an important question to evaluate the discriminative power of VA questions in assigning causes, and identify the questions that do no contribute to cause assignments across different domains. This can facilitate the design of shorter VA questionnaires in the future. The proposed framework can be extended to perform symptom selection by further shrinking the response probabilities across different causes of death to the same values.  
Third, 
while our model focuses on the prediction task, the discovery and inference of interpretable latent classes shared across different populations alone is a topic of significant practical interest \citep{de2019multi}. It is a challenging task in the VA context, as the collected symptoms could be measured inconsistently across domains and subject to reporting biases. This is related to the discussion of transportability in the causal inference literature \citep{wu2019causal,ackerman2019transportability}. 

Finally, we conclude by highlighting two additional open questions. 
First, the proposed model relies on reference deaths with known causes of death from multiple training domains. A common practice of collecting labeled VA data is by having physician experts read and assign a cause best describing the death. The assigned causes of death are treated as the ground truth. However, many factors could contribute to the misclassification of deaths and lead to noise in these training labels. In particular, when a new disease emerges and limited information about the disease is known, such as during the early periods of the COVID-19 pandemic, higher rate of misclassification may be expected from the training labels. A key challenge, therefore, will be to identify latent classes that correspond to misclassified or unseen causes. Methods dealing with label noises in classification tasks \citep{liu2020peer,liu2021understanding} could potentially improve cause-of-death assignment models and produce more realistic characterization of uncertainties.
Second,
a grand challenge with VA is to embed the analysis framework into the mortality surveillance system to detect changes and abnormalities. Modeling the temporal drift of the data distribution is highly useful for surveillance purposes.
Covariate-dependent modeling \citep{moran2021bayesian} is also important in this context, as additional information from both the population mortality surveillance and individual deaths could further help understand transferability and explain changes in mortality profile. We leave these topics for future work.

\bibliographystyle{apalike}
\bibliography{factor}

\clearpage
\appendix

\section{Sampler for the single-domain model}

\begin{enumerate}
\item Sample the cause-of-death assignments $Y_i$ and latent class indicator $Z_i$. To facilitate easier computation, we augment the data with latent indicator $\tilde D_i \in \{1, ..., G\}$ for observations from the target population and sample all latent indicators jointly with 
\[
  p(Y_i  = c, Z_i = k | X_i, D_i = 0, \blambda, \btheta) 
  \propto 
  \pi_c^{(0)}  \lambda_{ck} \prod_j (\theta_{ckj})^{x_{ij}}(1-\theta_{ckj})^{1-x_{ij}}.
\]
 \item For the case with new mixing weights, we need to sample the target-specific $\blambda^{(0)}$, which is the same as Step 2 of the training stage model for $g = 0$. 

  \item Update the CSMF vector in the target domain $\bpi^{(0)}|Y$ with
  \[
     \bpi^{(0)} |Y \sim \mbox{Dirichlet}(\alpha_{\pi, 1}^{(0)} + n^{(0)}_1, ..., \alpha_{\pi, C}^{(0)} + n^{(0)}_C)
  \]
  where $n^{(0)}_c = \sum_{i} \bm 1_{Y_i = c, D_i = 0}$.
 
\end{enumerate}

\section{Synthetic data simulation}
\label{sec:results-simulation}
We simulate deaths from $5$ training domains with known causes-of-death and one target domain without labeled data. We let $C = 20$ and $S = 50$ and $2,000$ deaths in each domain. We consider two distributions of causes in the training domains: (1) balanced cause-of-death distribution with $\pi^{(g)} \sim \mbox{Dirichlet}(5)$ and (2) unbalanced cause-of-death distribution with with $\pi^{(g)} \sim \mbox{Dirichlet}(0.5)$  
In each case, we consider four different data generating processes for $\blambda^{(g)}$:
\begin{enumerate}
  \item \textbf{Single domain with conditional independent symptoms}: We let $K = 1$ and $\lambda_{c1}^{(g)} = 1$. That is, we assume the distribution of $p(X, Y)$ is the same in all the five domains and that the symptoms are conditionally independent given the causes;

  \item \textbf{Single domain}:  We let $\blambda^{(g)} = \blambda$ for all sites with $K = 10$ latent classes and generate $\blambda_c \sim \mbox{Dirichlet}(1)$ for each cause of death independently;

  \item \textbf{Independent domains}:  We let $\blambda^{(g)}_{c} \sim_{iid} \mbox{Dirichlet}(0.1)$ for all training and target domains. The Dirichlet prior encourages the weights to concentrate on a small number of latent classes. The target domain could contain latent classes that are rarely represented in the training domains in this case;

  \item \textbf{Related domains}: We consider the case that two of the training domains are related to the target domain. More specifically, let $\blambda^{(g)}_{c} \sim_{iid} \mbox{Dirichlet}(0.1)$ for all training domains. For the target domain we generate $\blambda^{(0)}$ as the average of two randomly picked training sites with $\blambda^{(0)}_{c} = 0.5\blambda^{(g_{1})}_{c} + 0.5 \blambda^{(g_{2})}_{c}$ for all $c$.

  \item \textbf{Related domains with cause-specific association}: We consider the case where the training domains are related to the target domain and such association vary across causes of death. More specifically, let $\blambda^{(g)}_{c} \sim_{iid} \mbox{Dirichlet}(0.1)$ for all training domains, and $\blambda^{(0)}_{c} = \sum_g \eta_{cg}\blambda^{(g)}_c$ where $\bmeta_c \sim \mbox{Dirichlet}(0.5)$.

\end{enumerate} 

Since in the real data we expect the distribution of symptoms to be bimodal with heavier concentration around $0$, we generated the response probabilities independently from a mixture distribution $\theta_{ckj} \sim 0.9\mbox{Beta}(1, 5) + 0.1\mbox{Beta}(5, 1)$. We further randomly mask $30\%$ of the symptoms as missing. 

In each scenario, we fit the four versions of the proposed latent class model with $K = 10$ using the same setup of MCMC iterations as described in the main paper for the PHMRC data experiments.  To benchmark the performance with VA algorithms used in practice, we compare these metrics with InSilicoVA, which assumes symptoms are conditionally independent.  

\begin{figure}[!htb]
\includegraphics[width = \textwidth]{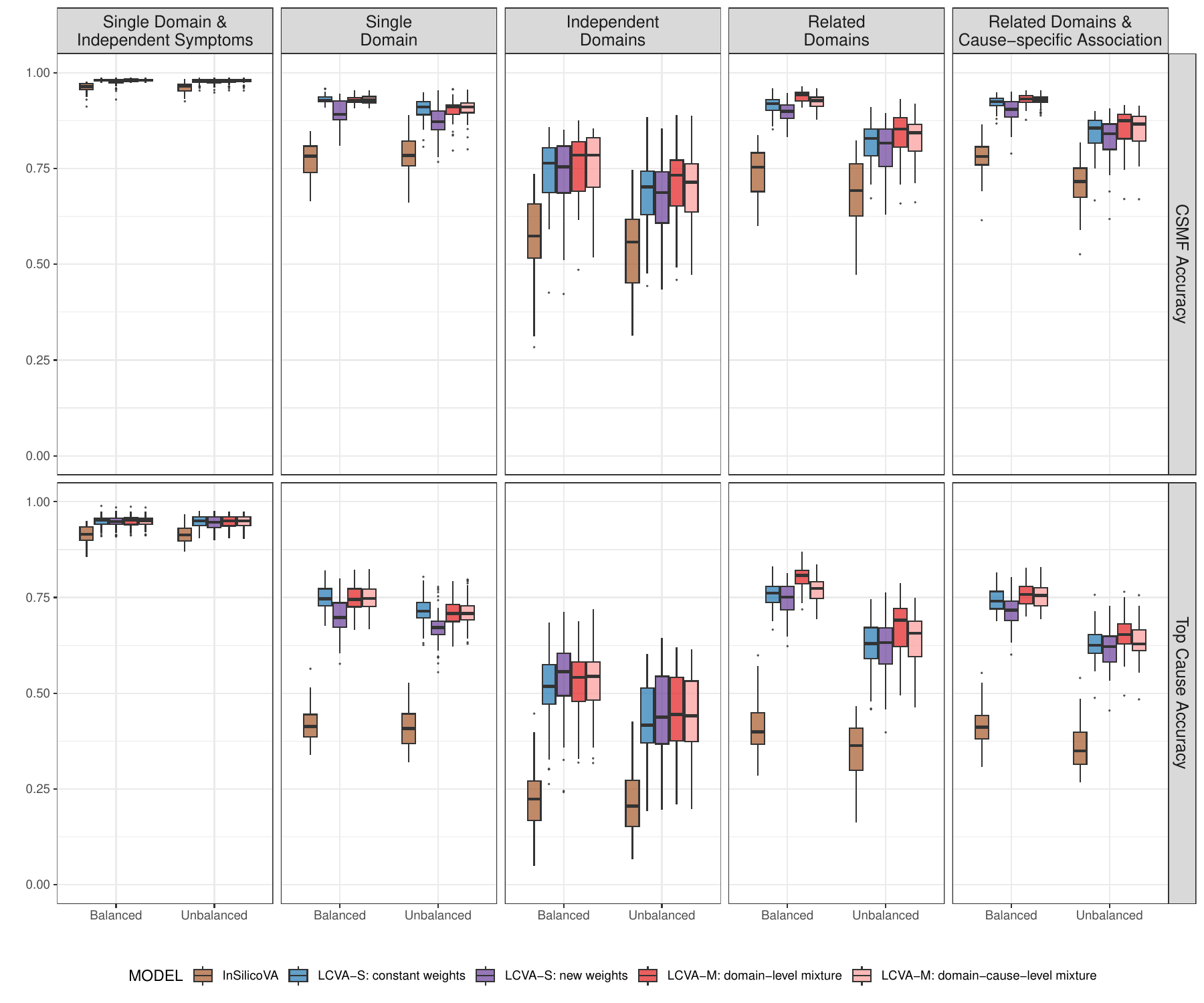}
\caption{CSMF accuracy (top row) and individual cause-of-death classification accuracy (bottom row) for the four simulation scenarios. \nameholder{} performs significantly better than InSilicoVA. \nameholder{}-S with constant weights achieves the best performance when data come from a single domain and \nameholder{}-M with domain-level mixture achieves the best performance in the three cases with multiple domains. }
\label{Sfig:sim}
\end{figure}

Figure \ref{Sfig:sim} shows the performance of the algorithms. For case (1), all models achieves almost perfect prediction. The performance of  InSilicoVA drops significantly in the latter four cases where the symptoms are dependent. Overall, the performance of the four variations of \nameholder{} are consistently better than InSilicoVA.  Within the four parameterizations of \nameholder{}, the single-domain model with constant weights achieves the best results when data are truly generated from the same domain without heterogeneity. Multi-domain model with domain-level mixture achieves the best performance in case (4) where the target domain consists of mixtures of latent classes from two existing domains, with overall high prediction accuracy in both metrics across all experiments.  

\section{Additional results on the PHMRC dataset}

In this section, we present additional details and results for the analysis of the PHMRC dataset discussed in the main manuscript. 

\subsection{Site-specific CSMFs in the PHMRC dataset}

\begin{figure}[!ht]
\centering
\includegraphics[width = .8\textwidth]{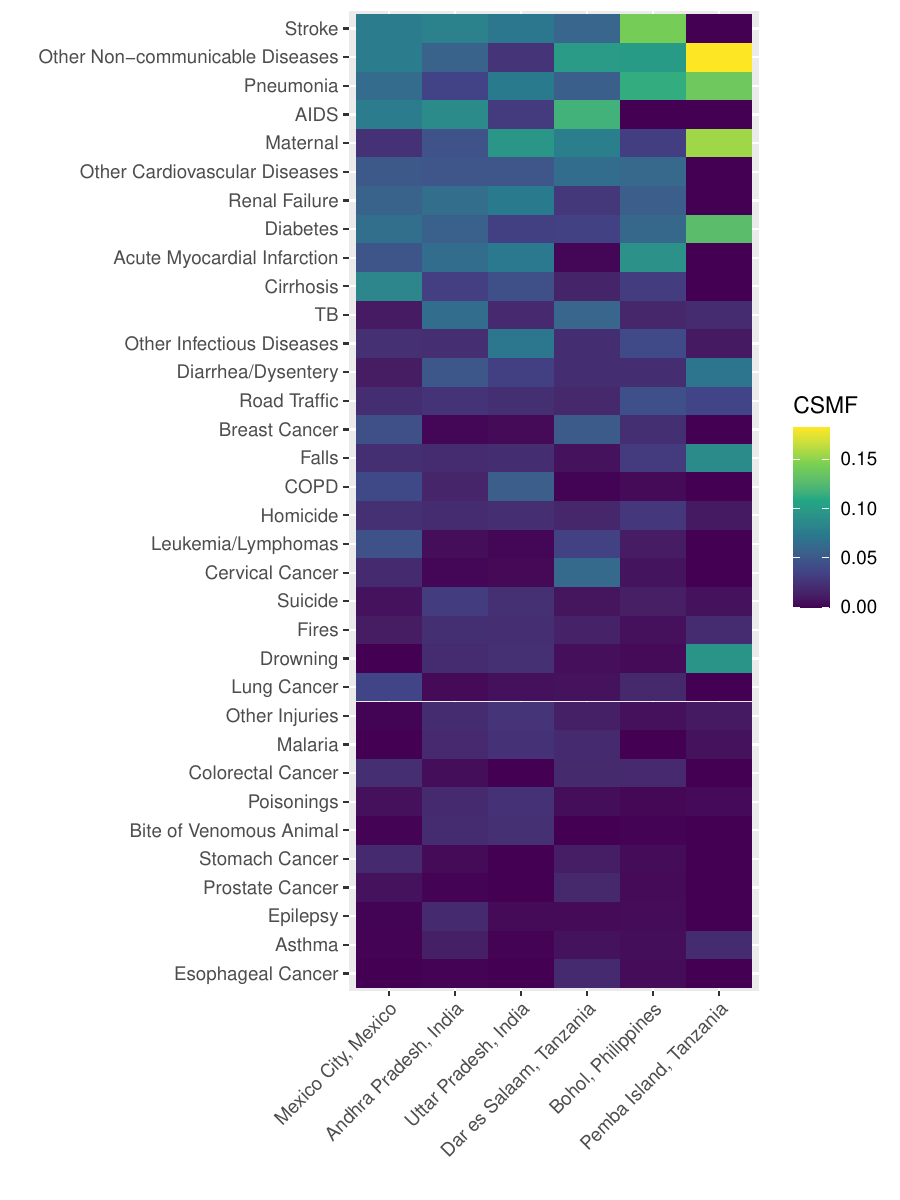}
\caption{True CSMFs in the six study sites in the PHMRC data.}
\label{Sfig:csmf-emp}
\end{figure}

Figure \ref{Sfig:csmf-emp} shows the empirical distribution of the causes in the six study sites in the PHMRC data, which we treat as the true CSMFs in our evaluation.

\subsection{Sensitivity analysis of $K$ on the leave-one-site-out experiment}

\begin{figure}[htb]
  \includegraphics[width = \textwidth]{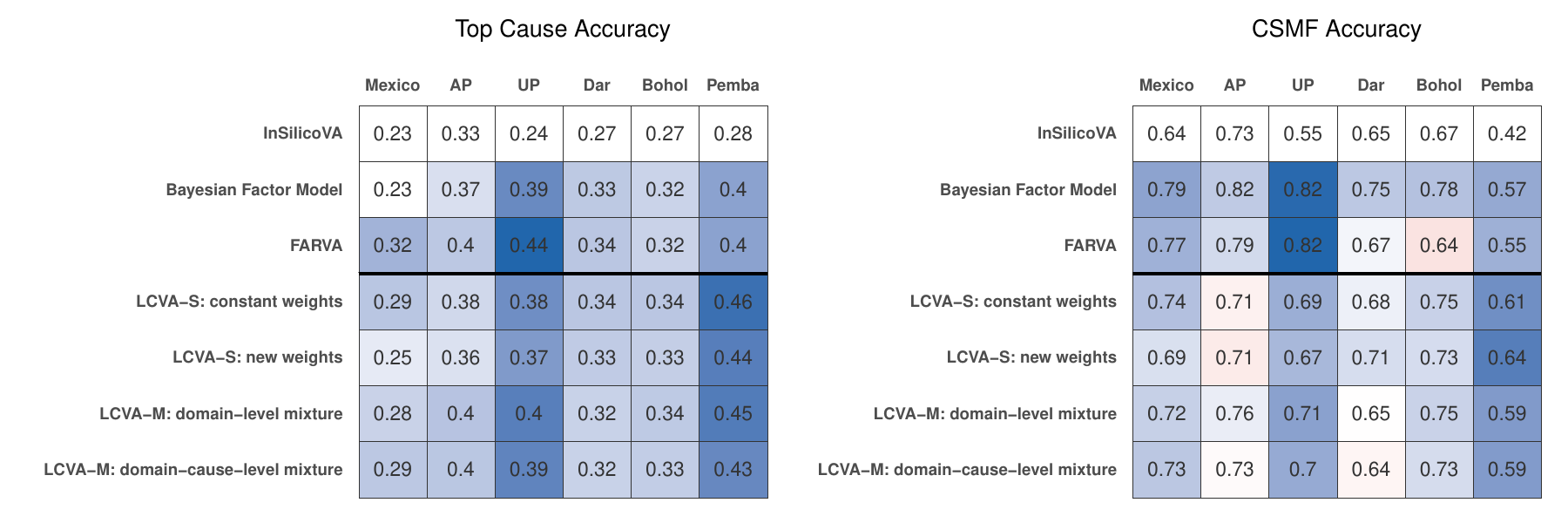}
  \caption{Individual-level top cause prediction accuracy (left) and CSMF accuracy (right) of different models. The proposed models are fitted with $\bm{K = 3}$. The cells are colored by the relative difference from the InSilicoVA results, where blue cells indicate higher accuracy and red cells indicate lower accuracy. All variations of \nameholder{} demonstrate good performance that is significantly higher than InSilicoVA except in one experiment. Bayesian factor model and FARVA also show comparable improvements.}
  \label{Sfig:phmrc-original-1}
  \end{figure}

\begin{figure}[htb]
  \includegraphics[width = \textwidth]{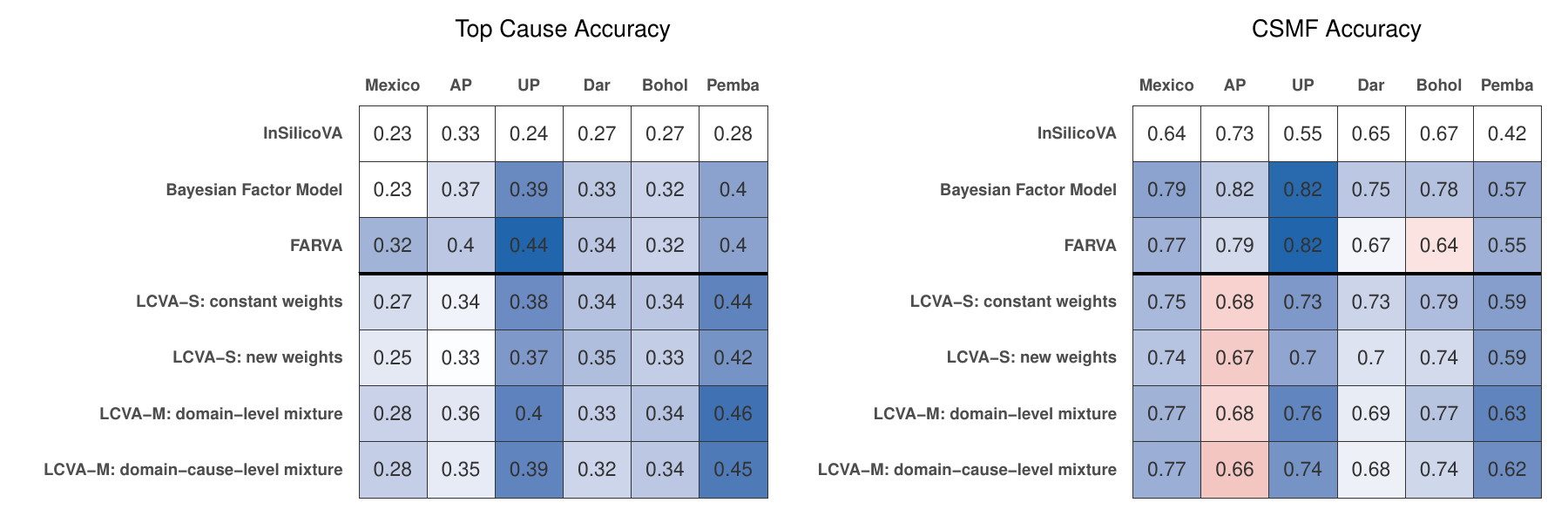}
  \caption{Individual-level top cause prediction accuracy (left) and CSMF accuracy (right) of different models. The proposed models are fitted with $\bm{K = 15}$. The cells are colored by the relative difference from the InSilicoVA results, where blue cells indicate higher accuracy and red cells indicate lower accuracy. All variations of \nameholder{} demonstrate good performance that is significantly higher than InSilicoVA except in one experiment. Bayesian factor model and FARVA also show comparable improvements.}
  \label{Sfig:phmrc-original-3}
  \end{figure}

Figure \ref{Sfig:phmrc-original-1} to \ref{Sfig:phmrc-original-3}  summarize the top cause accuracy and CSMF accuracy for the different models evaluated on the six sites when $K = 3, 4, 5,$ and $15$ respectively. The results are similar to the figure presented in the main paper with $K = 10$.

  \begin{figure}[htb]
  \includegraphics[width = \textwidth]{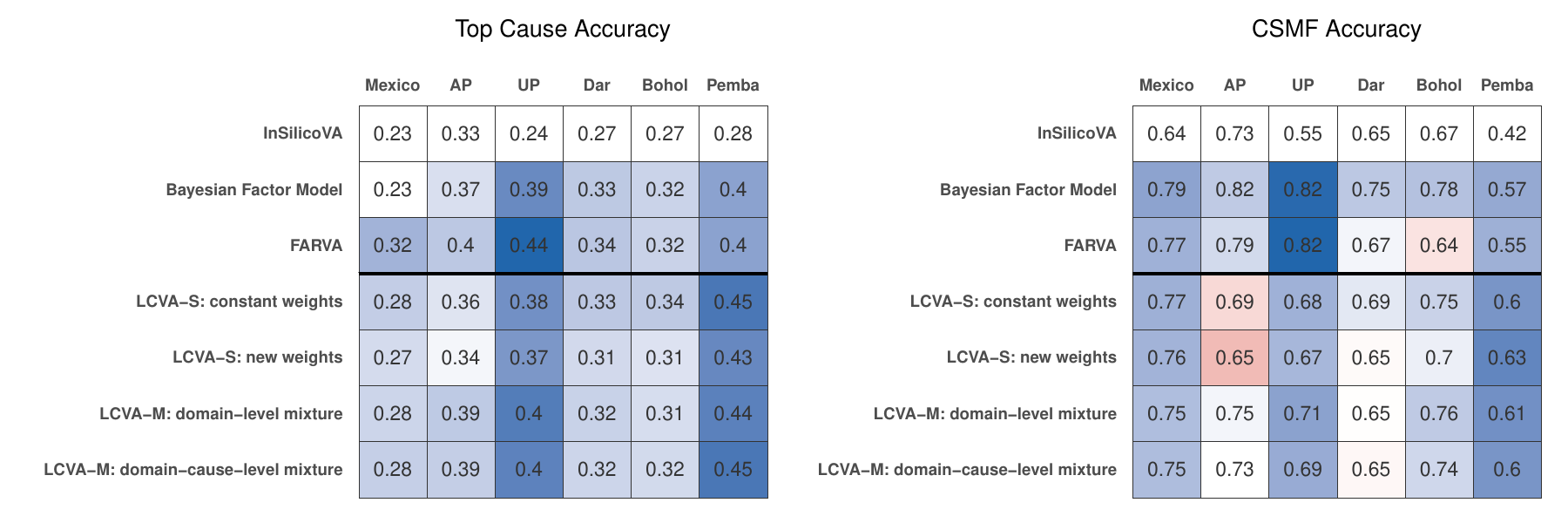}
  \caption{Individual-level top cause prediction accuracy (left) and CSMF accuracy (right) of different models. The proposed models are fitted with $\bm{K = 4}$. The cells are colored by the relative difference from the InSilicoVA results, where blue cells indicate higher accuracy and red cells indicate lower accuracy. All variations of \nameholder{} demonstrate good performance that is significantly higher than InSilicoVA except in one experiment. Bayesian factor model and FARVA also show comparable improvements.}
  \label{Sfig:phmrc-original-2}
  \end{figure}
    \begin{figure}[htb]
  \includegraphics[width = \textwidth]{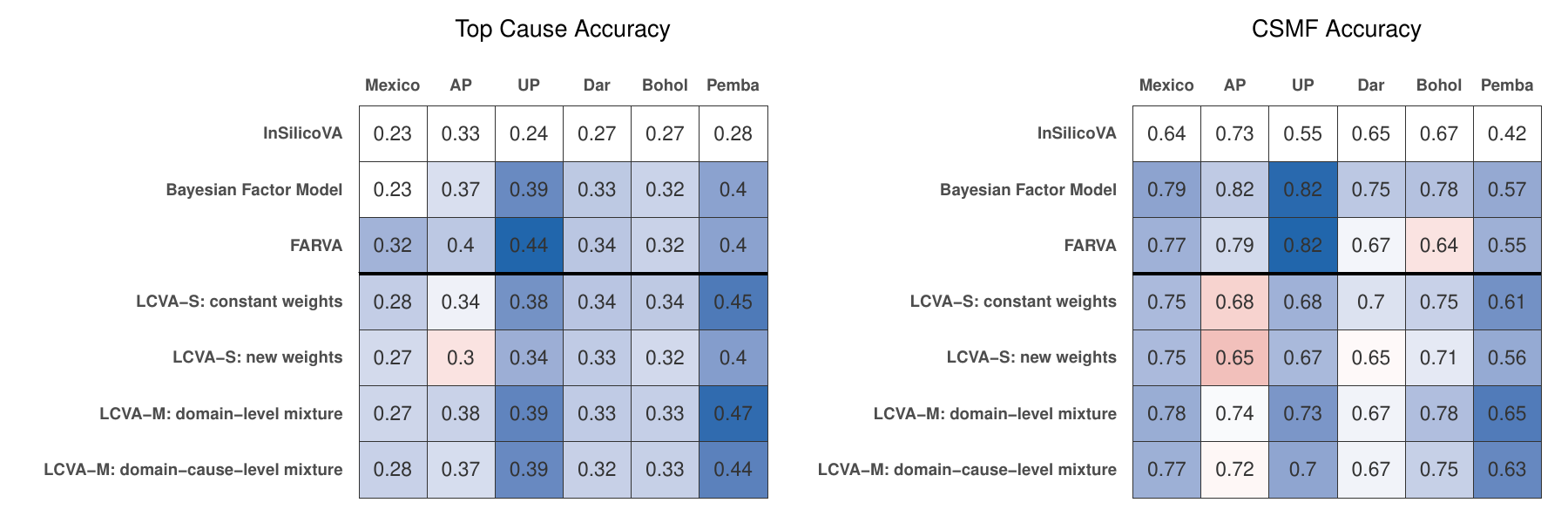}
  \caption{Individual-level top cause prediction accuracy (left) and CSMF accuracy (right) of different models. The proposed models are fitted with $\bm{K = 5}$. The cells are colored by the relative difference from the InSilicoVA results, where blue cells indicate higher accuracy and red cells indicate lower accuracy. All variations of \nameholder{} demonstrate good performance that is significantly higher than InSilicoVA except in one experiment. Bayesian factor model and FARVA also show comparable improvements.}
  \label{Sfig:phmrc-original-2}
  \end{figure}

\clearpage
\subsection{Additional results on resampled PHMRC experiment}

\begin{figure}[htb]
  \includegraphics[width = \textwidth]{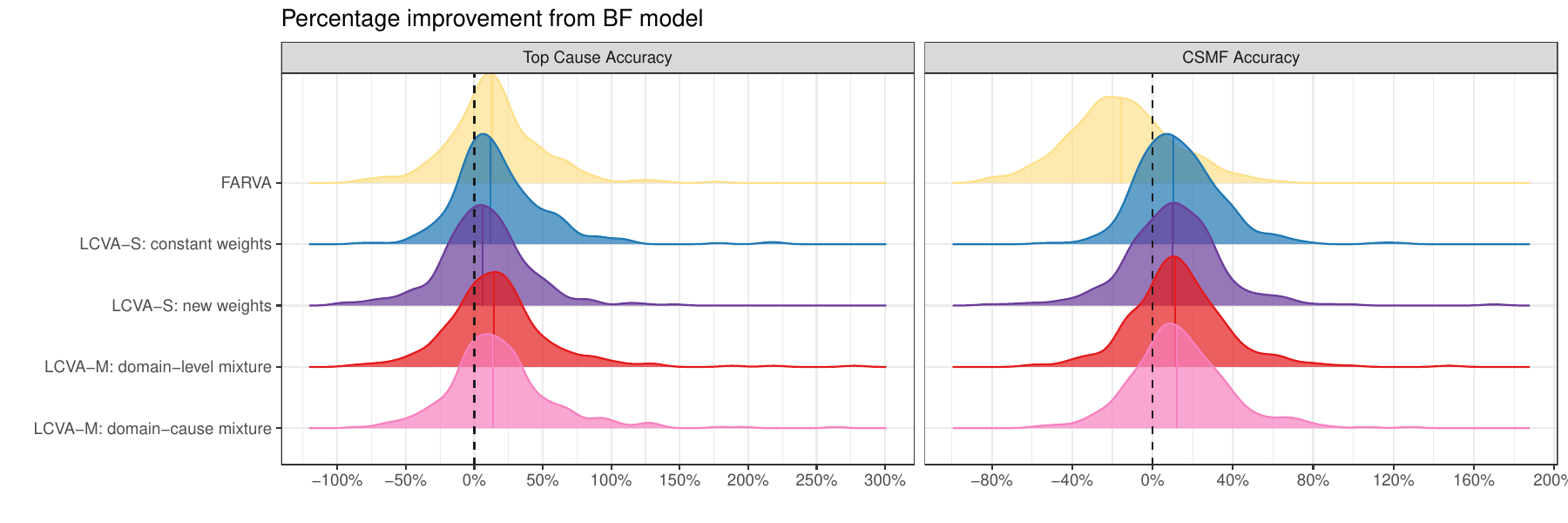}
  \caption{Distribution of \textbf{relative} improvement over the Bayesian factor model, i.e., $(Acc_m - Acc_{BF}) / ACC_{BF}$, for each model $m$ in terms of the top cause accuracy (left) and CSMF accuracy (right) by different models on all $300$ resampled target datasets. }
  \label{Sfig:phmrc-resample-4}
  \end{figure}

In this subsection, we first provide additional summaries for the relative improvement over the Bayesian factor model, in addition to the results presented in the main paper. The mean, and the $25$th and $75$th percentiles of the relative improvements of all models over the Bayesian factor model are summarized in Table \ref{Stab:pdiff}. Except for the top cause accuracy in Dar es Salaam, the two versiosns of LCVA-M improves from Bayesian factor models in all other cases, and the improvements are quite dramatic in Mexico city, Andhra Pradesh, Uttar Pradesh, and Bohol.  Figure \ref{Sfig:phmrc-resample-4} show the distribution of relative improvements over all $300$ resampled datasets for each model. Across all experiments, the average improvements of \nameholder{}-M are similar. Take \nameholder{}-M with domain-cause-level mixture for example, we observe on average around $19\%$ and $14\%$ improvement in terms of the top cause accuracy and CSMF accuracy respectively.

  \begin{table}
  \renewcommand{\arraystretch}{0.5}
 \begin{tabular}{llll}
  \hline
Site & Model & Top Cause Accuracy & CSMF Accuracy \\
  \hline
Mexico & FARVA & 47\% (21\%, 66\%) & -22\% (-41\%, -7\%) \\
   & LCVA-S: constant weights & 44\% (12\%, 62\%) & 20\% (9\%, 31\%) \\
   & LCVA-S: new weights & 24\% (9\%, 38\%) & 15\% (3\%, 25\%) \\
   & LCVA-M: domain-level mixture & 46\% (19\%, 55\%) & 20\% (6\%, 32\%) \\
   & LCVA-M: domain-cause mixture & 53\% (23\%, 64\%) & 23\% (6\%, 34\%) \\ \hline
  AP & FARVA & 17\% (4\%, 22\%) & -13\% (-30\%, 0\%) \\
   & LCVA-S: constant weights & 11\% (-2\%, 21\%) & 8\% (-3\%, 20\%) \\
   & LCVA-S: new weights & 4\% (-10\%, 10\%) & 4\% (-8\%, 18\%) \\
   & LCVA-M: domain-level mixture & 12\% (0\%, 25\%) & 8\% (-1\%, 20\%) \\
   & LCVA-M: domain-cause mixture & 13\% (1\%, 26\%) & 8\% (-3\%, 23\%) \\\hline
  UP & FARVA & 22\% (8\%, 37\%) & 0\% (-18\%, 18\%) \\
   & LCVA-S: constant weights & 11\% (0\%, 21\%) & 15\% (3\%, 25\%) \\
   & LCVA-S: new weights & 3\% (-14\%, 12\%) & 11\% (-3\%, 22\%) \\
   & LCVA-M: domain-level mixture & 17\% (4\%, 26\%) & 19\% (7\%, 26\%) \\
   & LCVA-M: domain-cause mixture & 16\% (3\%, 28\%) & 19\% (6\%, 30\%) \\\hline
  Dar & FARVA & 3\% (-9\%, 19\%) & -24\% (-39\%, -11\%) \\
   & LCVA-S: constant weights & 6\% (-5\%, 14\%) & 8\% (-3\%, 19\%) \\
   & LCVA-S: new weights & 3\% (-10\%, 16\%) & 13\% (4\%, 22\%) \\
   & LCVA-M: domain-level mixture & -5\% (-22\%, 9\%) & 1\% (-10\%, 11\%) \\
   & LCVA-M: domain-cause mixture & -5\% (-19\%, 8\%) & 2\% (-10\%, 10\%) \\\hline
  Bohol & FARVA & 20\% (-3\%, 45\%) & -13\% (-30\%, 1\%) \\
   & LCVA-S: constant weights & 29\% (8\%, 56\%) & 22\% (6\%, 38\%) \\
   & LCVA-S: new weights & 19\% (2\%, 35\%) & 20\% (4\%, 32\%) \\
   & LCVA-M: domain-level mixture & 29\% (-1\%, 56\%) & 24\% (7\%, 39\%) \\
   & LCVA-M: domain-cause mixture & 30\% (-5\%, 59\%) & 25\% (6\%, 39\%) \\\hline
  Pemba & FARVA & -20\% (-36\%, 4\%) & -22\% (-42\%, 1\%) \\
   & LCVA-S: constant weights & 9\% (-11\%, 26\%) & 6\% (-11\%, 16\%) \\
   & LCVA-S: new weights & -11\% (-42\%, 13\%) & 1\% (-25\%, 15\%) \\
   & LCVA-M: domain-level mixture & 4\% (-26\%, 25\%) & 6\% (-12\%, 18\%) \\
   & LCVA-M: domain-cause mixture & 7\% (-18\%, 25\%) & 6\% (-9\%, 19\%) \\
   \hline
\end{tabular}
\caption{Mean \textbf{relative} improvement compared to Bayesian Factor model, i.e., $(Acc_m - Acc_{BF}) / ACC_{BF}$, for for each model $m$ in terms of the top cause accuracy and CSMF accuracy. The intervals are $25$th and $75$th percentiles of the improvement. }
  \label{Stab:pdiff}
  \end{table}

\begin{figure}[!htb]
  \includegraphics[width = \textwidth]{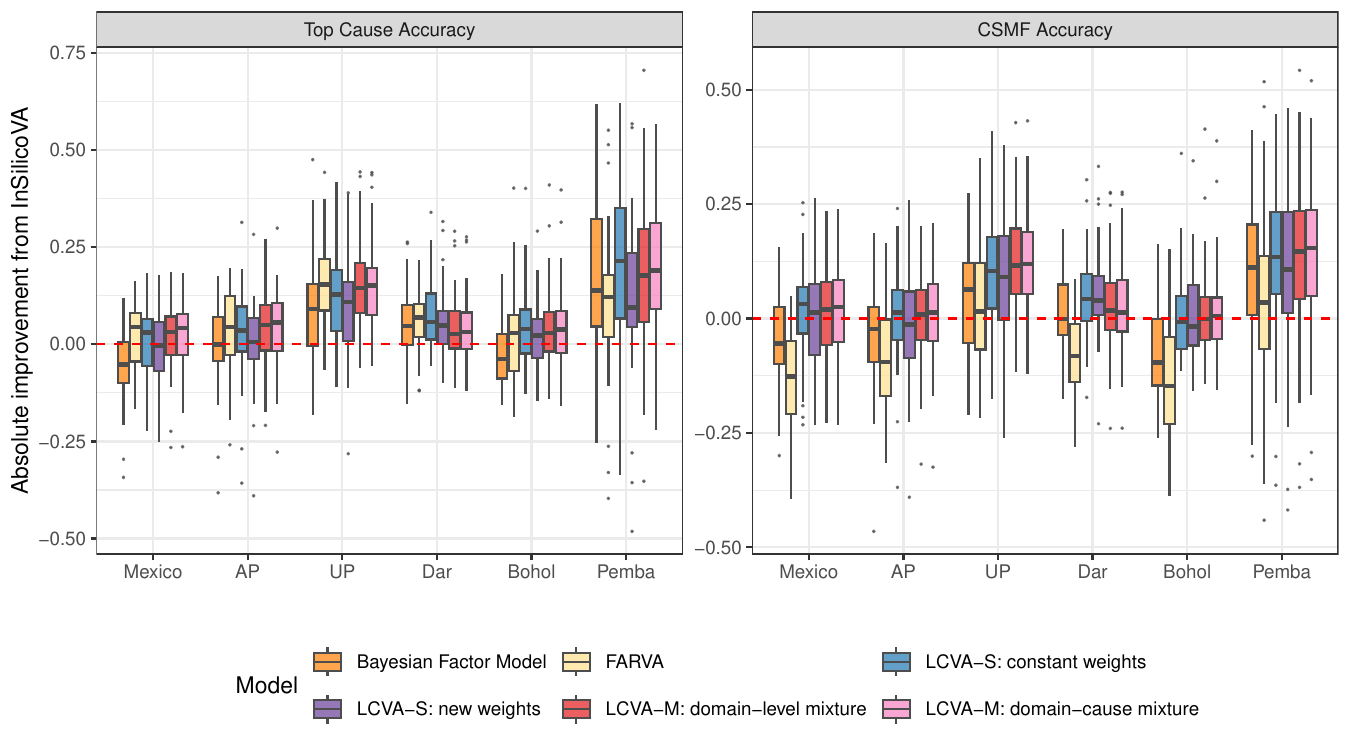}
  \caption{Boxplot of \textbf{absolute} improvement over InSilicoVA \citep{mccormick2016probabilistic} in top cause accuracy (left) and CSMF accuracy (right) by different models on $50$ resampled target domain datasets based on deaths from each of the six sites.}
  \label{Sfig:phmrc-resample-1}
  \end{figure}

In addition to the comparison with Bayesian factor model, Figure \ref{Sfig:phmrc-resample-1} show the relative improvements from InSilicoVA computed as $(Acc_m - Acc_{ins})$ for all models considered. We did not compute the percentage improvement because the accuracy metrics of InSilicoVA on many resampled datasets are too low, which makes the percentage improvement very high and not easily readable.

\section{Additional results when limited target data is available}

In this section, we present additional results on the analysis of calibrating \nameholder{} and InSilicoVA with additional labeled data from the target domain. 

\subsection{Cause of death aggregation}
Table \ref{Stab:agg} shows the three levels of the cause list we use in the calibration process. 
The algorithms are fitted on the original $34$ cause and the individual level predictions are aggregated to the other two broader categories for calibration.
The result with the $34$-cause list is presented in the main manuscript.

\begin{table}[!ht]
\renewcommand{\arraystretch}{0.5}
\centering
\begin{tabular}{lll}
  \hline
34 Causes & 11 Causes & 5 Causes \\
  \hline
Acute Myocardial Infarction & Disease of the Circulatory System & Circulatory \\
  Other Cardiovascular Diseases & Disease of the Circulatory System & Circulatory \\
  Stroke & Disease of the Circulatory System & Circulatory \\
  Bite of Venomous Animal & External & External \\
  Drowning & External & External \\
  Falls & External & External \\
  Fires & External & External \\
  Homicide & External & External \\
  Other Injuries & External & External \\
  Poisonings & External & External \\
  Road Traffic & External & External \\
  Suicide & External & External \\
  AIDS & Infectious and Parasitic Diseases & Infectious \\
  Diarrhea/Dysentery & Infectious and Parasitic Diseases & Infectious \\
  Malaria & Infectious and Parasitic Diseases & Infectious \\
  Other Infectious Diseases & Infectious and Parasitic Diseases & Infectious \\
  Pneumonia & Infectious and Parasitic Diseases & Infectious \\
  TB & Infectious and Parasitic Diseases & Infectious \\
  Maternal & Maternal & Maternal \\
  Cirrhosis & Gastrointestinal disorders & Non-Communicable \\
  Epilepsy & Mental and Nervous System Disorders & Non-Communicable \\
  Breast Cancer & Neoplasms & Non-Communicable \\
  Cervical Cancer & Neoplasms & Non-Communicable \\
  Colorectal Cancer & Neoplasms & Non-Communicable \\
  Esophageal Cancer & Neoplasms & Non-Communicable \\
  Leukemia/Lymphomas & Neoplasms & Non-Communicable \\
  Lung Cancer & Neoplasms & Non-Communicable \\
  Prostate Cancer & Neoplasms & Non-Communicable \\
  Stomach Cancer & Neoplasms & Non-Communicable \\
  Diabetes & Nutritional and Endocrine Disorders  & Non-Communicable \\
  Other Non-communicable Diseases & Other Noncommunicable Diseases & Non-Communicable \\
  Renal Failure & Renal Disorders & Non-Communicable \\
  Asthma & Respiratory disorders & Non-Communicable \\
  COPD & Respiratory disorders & Non-Communicable \\
   \hline
\end{tabular}
\caption{Two levels of aggregation of the $34$ causes of death in the PHMRC dataset in the calibration analysis.}
\label{Stab:agg}
\end{table}

\subsection{Calibration on the aggregated cause list}

\begin{figure}[!ht]
\includegraphics[width = \textwidth]{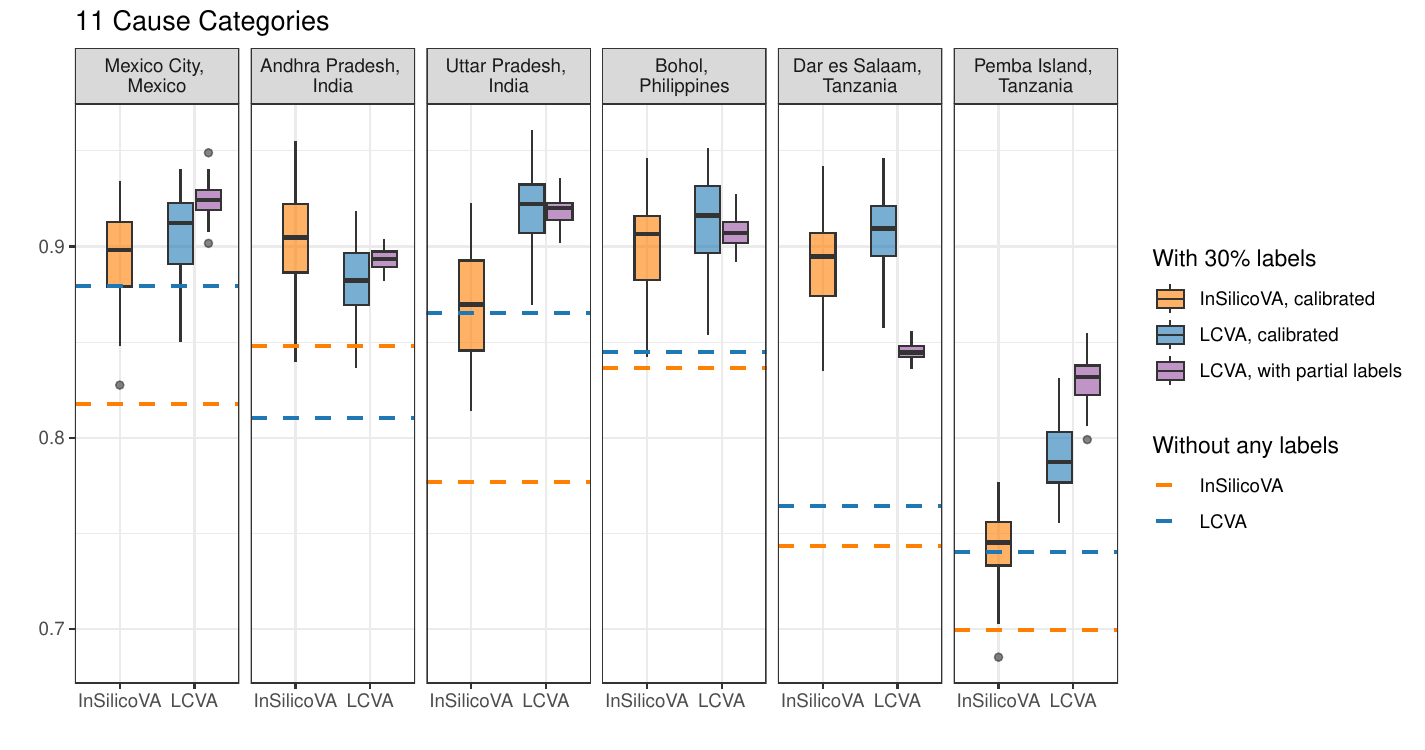}
\caption{Comparison of \nameholder{}-M with domain-level mixture and InSilicoVA when the cause of death of $30\%$ of data in the target domain are known. The causes of deaths have been aggregated into $11$ broad categories.  The horizontal line indicates the CSMF accuracy of the two methods before calibrating to the labeled deaths. The box plot shows the distribution of the calibrated CSMF accuracy over $50$ resampled labeled dataset.}
\label{Sfig:calib-11}
\end{figure}

\begin{figure}[!ht]
\includegraphics[width = \textwidth]{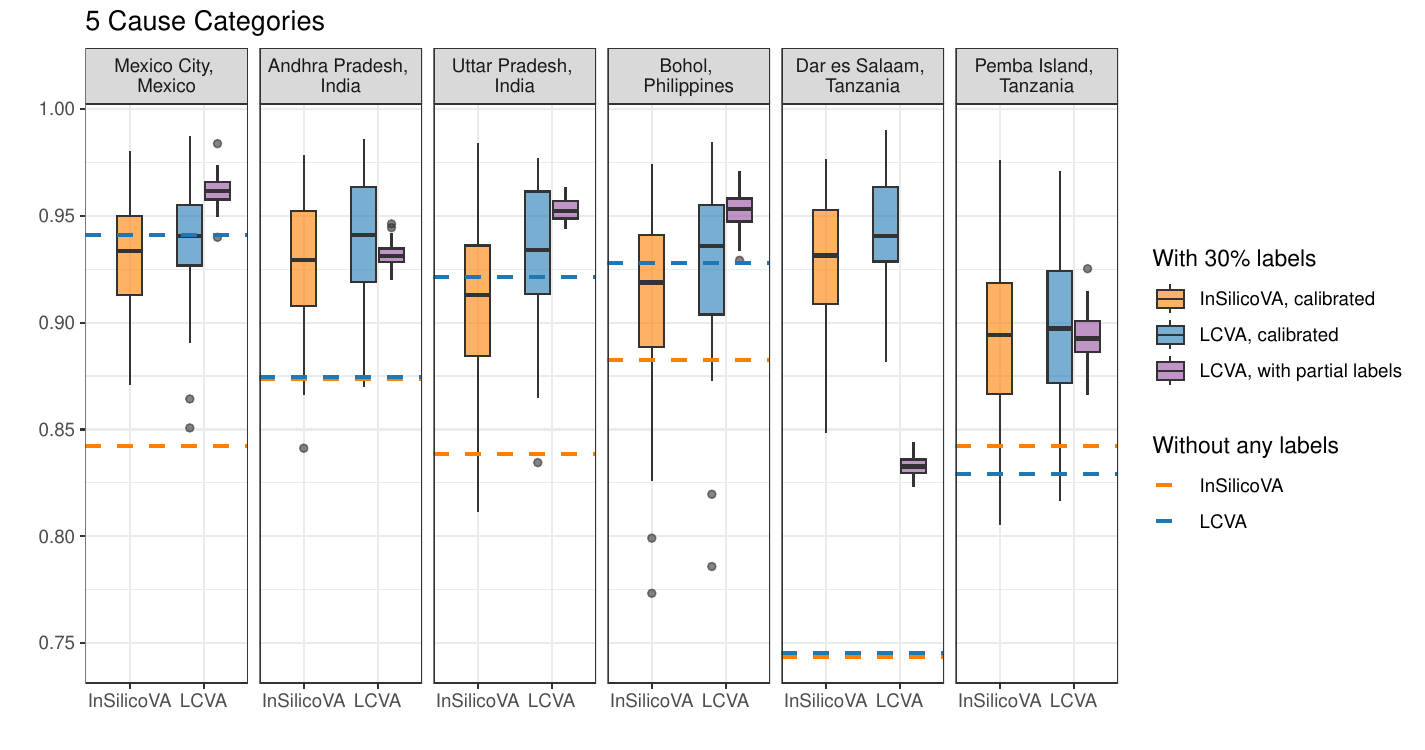}
\caption{Comparison of \nameholder{}-M with domain-level mixture and InSilicoVA when the cause of death of $30\%$ of data in the target domain are known. The causes of deaths have been aggregated into $5$ broad categories. The horizontal line indicates the CSMF accuracy of the two methods before calibrating to the labeled deaths. The box plot shows the distribution of the calibrated CSMF accuracy over $50$ resampled labeled dataset.}
\label{Sfig:calib-5}
\end{figure}

Figure \ref{Sfig:calib-11} compares the calibrated CSMF accuracy on the $11$-cause list. Except for Dar es Salaam, the joint modeling approach \nameholder{} with partial labels achieve comparable or higher CSMF accuracy than the calibration approach.  Figure \ref{Sfig:calib-5} compares the calibrated CSMF accuracy on the broader $5$-cause list, and the performance of the calibrated methods are similar. 
This analysis suggest that our approach to directly model the joint distribution of symptoms across domains is beneficial both with and without labeled data from the target domain, with the greatest improvements from the conditional independence model when the number of causes to consider is large.

\section{Convergence Analysis}
\begin{figure}[!ht]
\centering
\includegraphics[width = \textwidth]{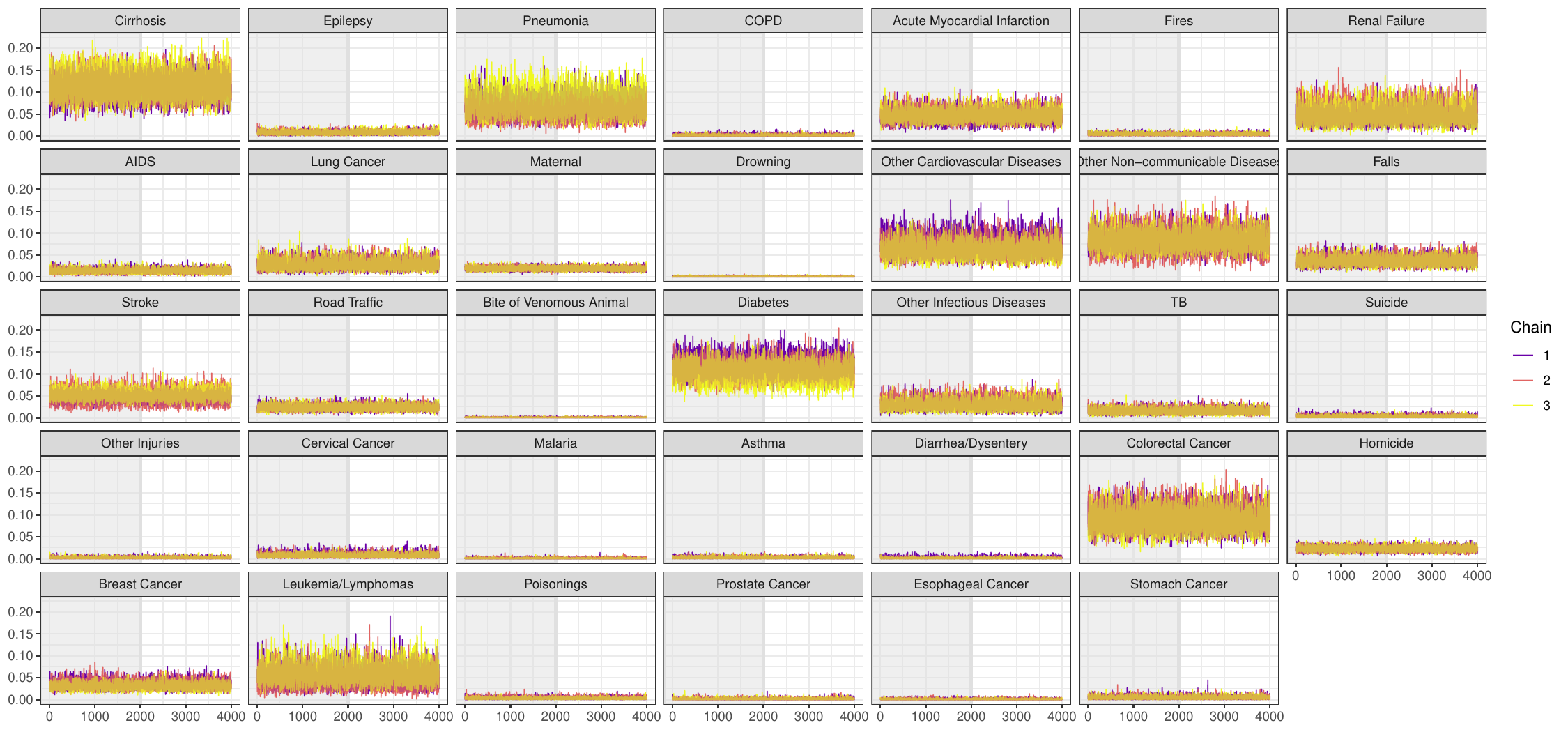}
\includegraphics[width = 0.8\textwidth]{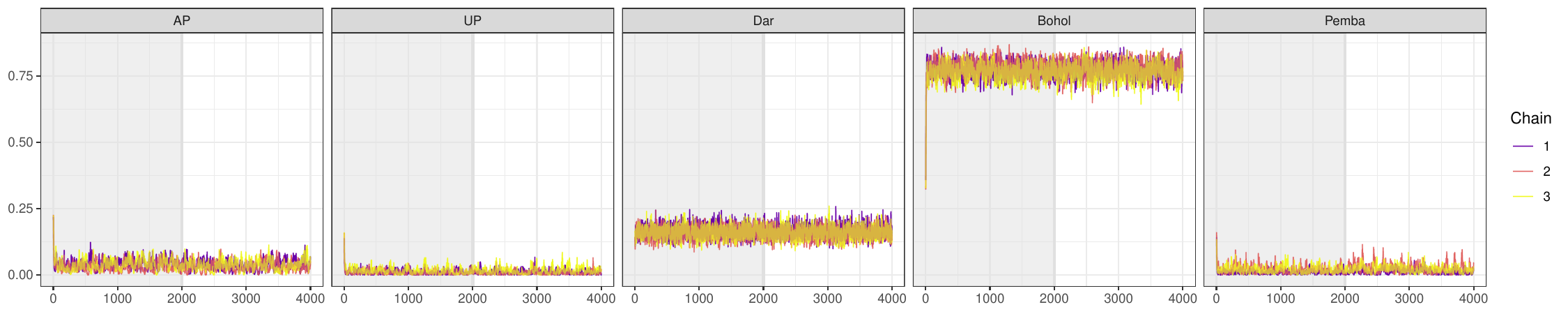}
\caption{Mexico City as target site: Trace plots for each CSMF, $\pi_c$ (top), and site similarity parameter $\eta_g$ (bottom) from three chains. The burn-in period is shaded in gray.}
\label{Sfig:conv-1}
\end{figure}

\begin{figure}[!ht]
\centering
\includegraphics[width = \textwidth]{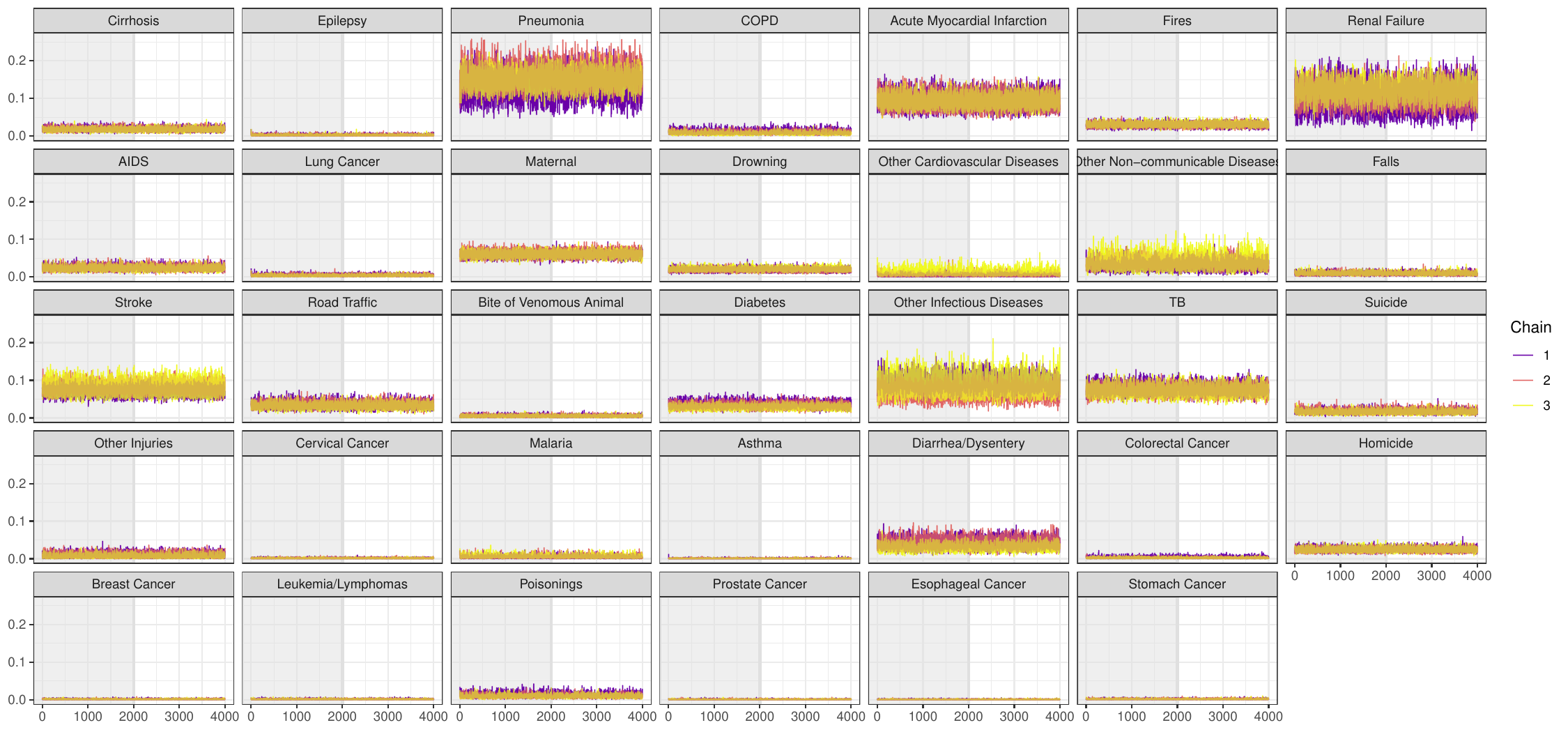}
\includegraphics[width = 0.8\textwidth]{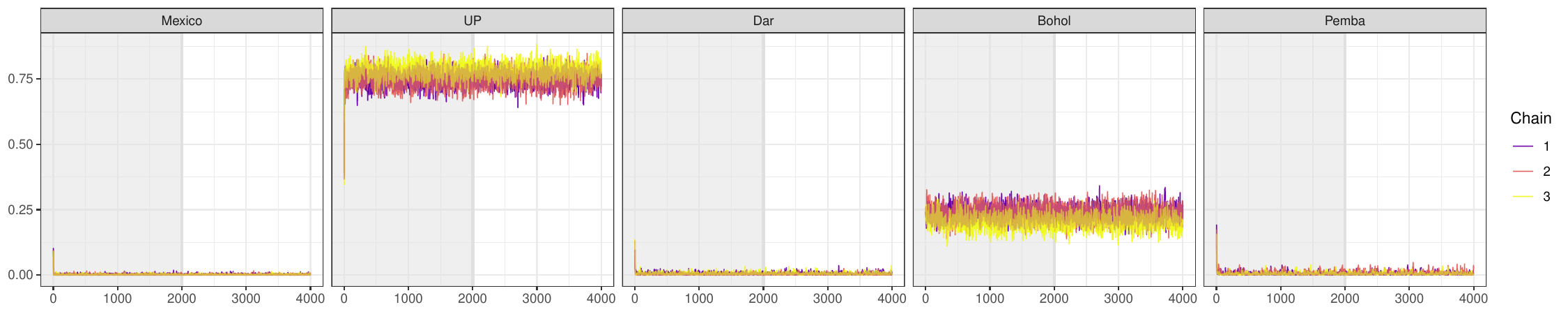}
\caption{Andhra Pradesh as target site: Trace plots for each CSMF, $\pi_c$ (top), and site similarity parameter $\eta_g$ (bottom) from three chains. The burn-in period is shaded in gray.}
\label{Sfig:conv-2}
\end{figure}

\begin{figure}[!ht]
\centering
\includegraphics[width = \textwidth]{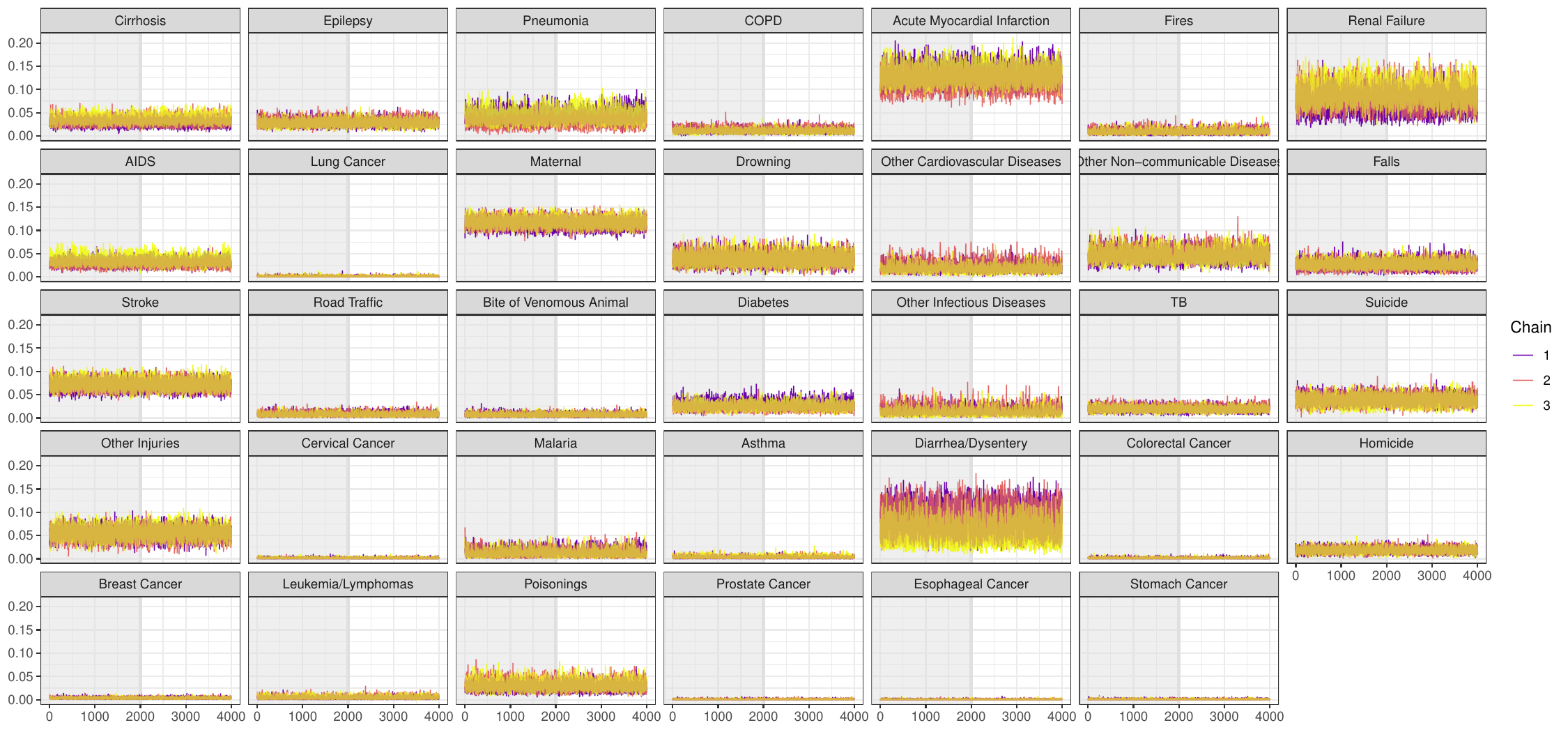}
\includegraphics[width = 0.8\textwidth]{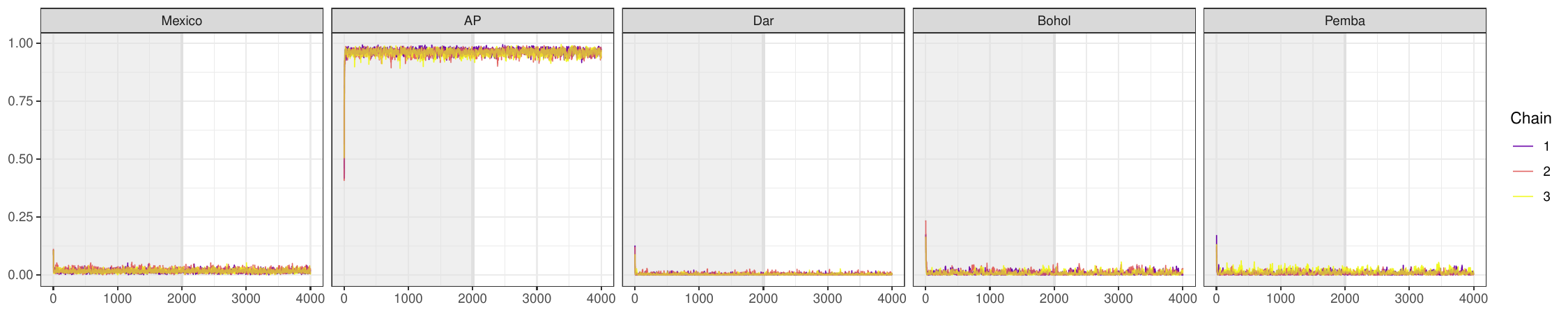}
\caption{Uttar Pradesh as target site: Trace plots for each CSMF, $\pi_c$ (top), and site similarity parameter $\eta_g$ (bottom) from three chains. The burn-in period is shaded in gray.}
\label{Sfig:conv-3}
\end{figure}

\begin{figure}[!ht]
\centering
\includegraphics[width = \textwidth]{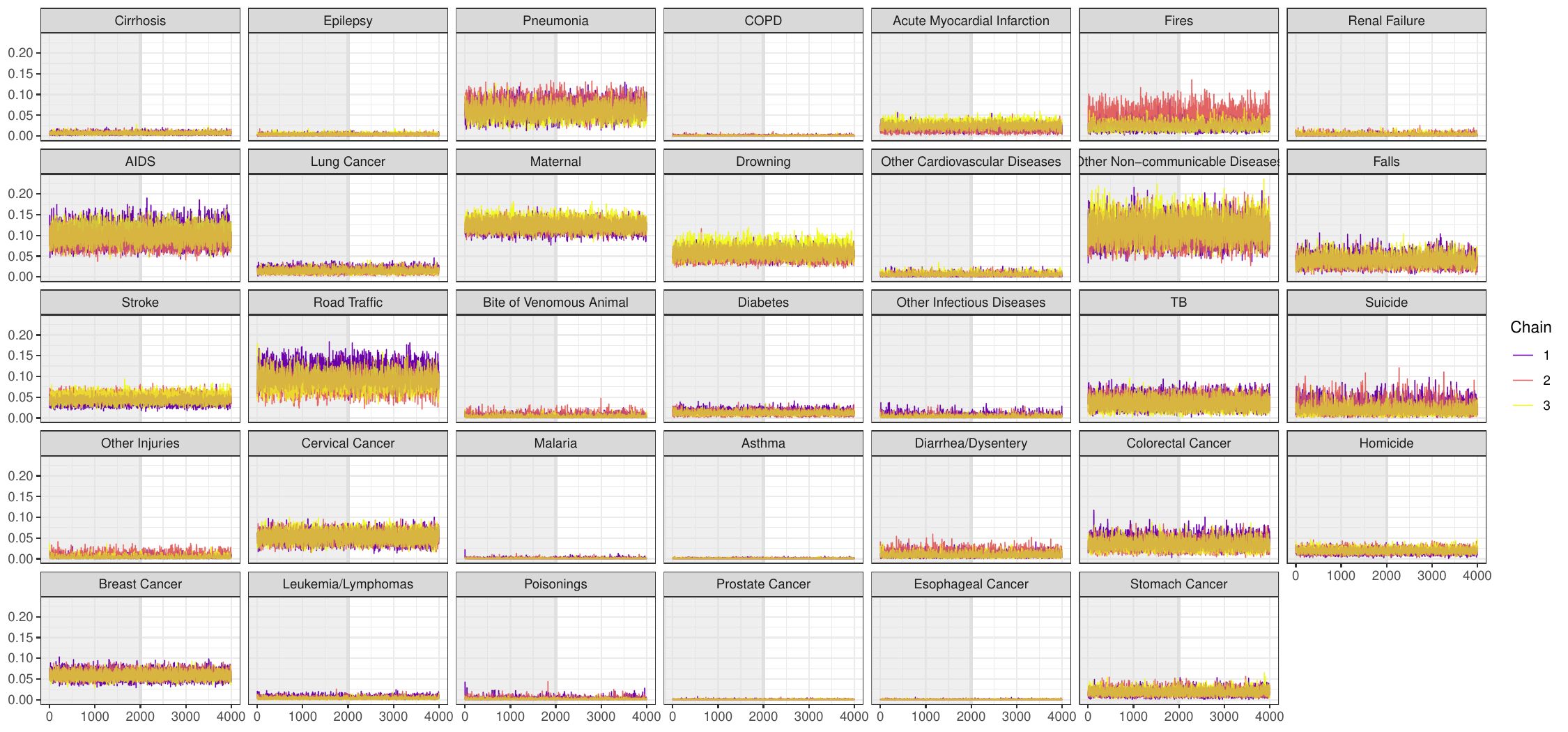}
\includegraphics[width = 0.8\textwidth]{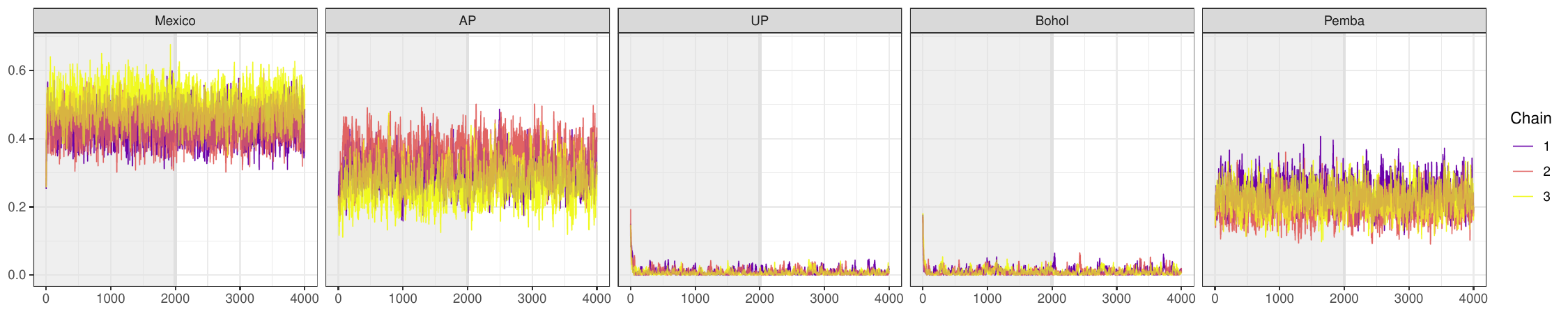}
\caption{Dar es Salaam as target site: Trace plots for each CSMF, $\pi_c$ (top), and site similarity parameter $\eta_g$ (bottom) from three chains. The burn-in period is shaded in gray.}
\label{Sfig:conv-4}
\end{figure}

\begin{figure}[!htb]
\centering
\includegraphics[width = \textwidth]{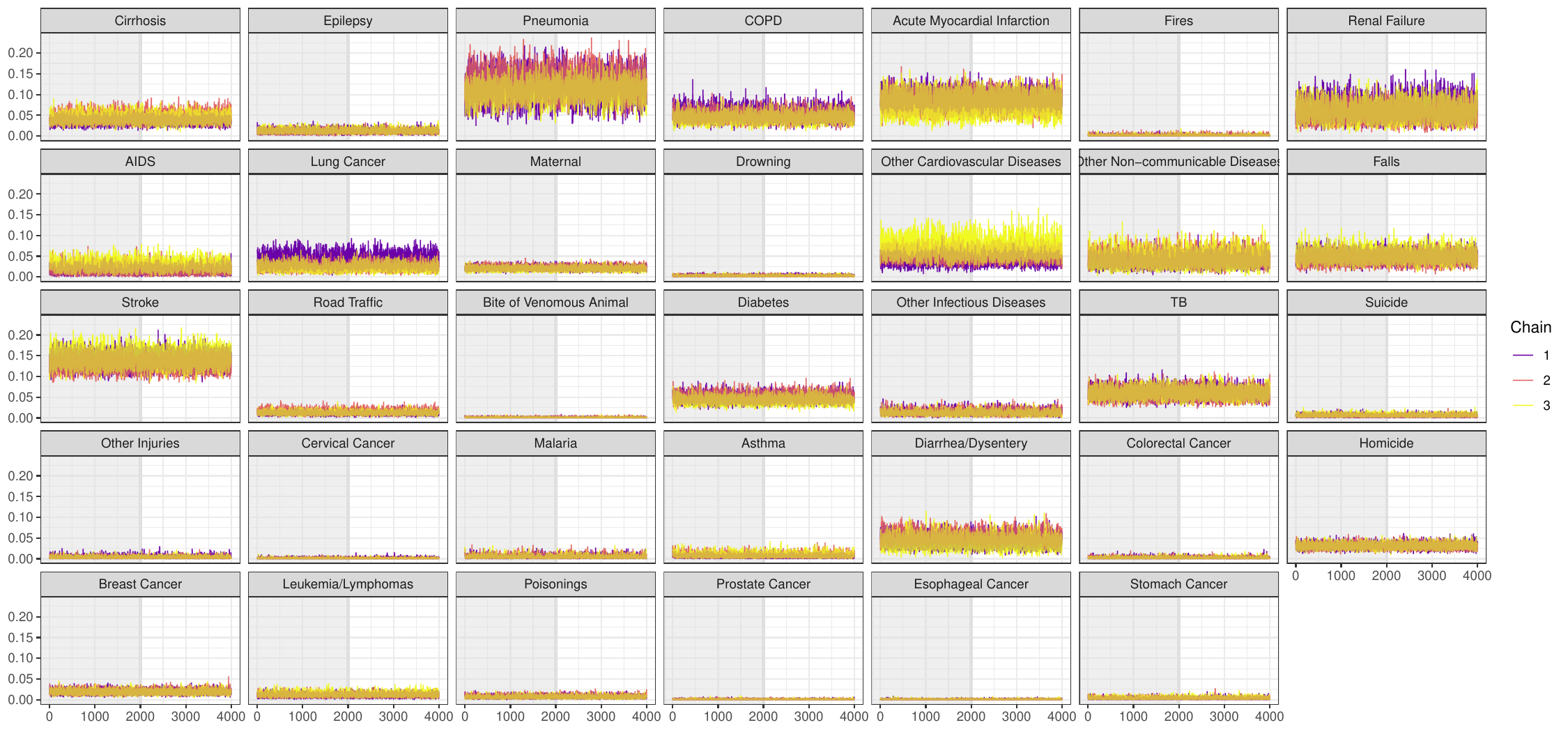}
\includegraphics[width = 0.8\textwidth]{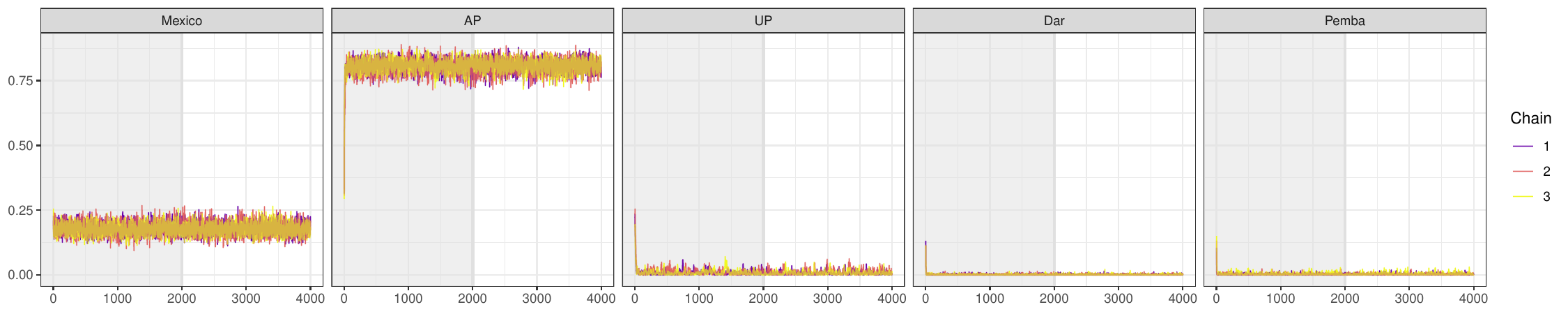}
\caption{Bohol as target site: Trace plots for each CSMF, $\pi_c$ (top), and site similarity parameter $\eta_g$ (bottom) from three chains. The burn-in period is shaded in gray.}
\label{Sfig:conv-5}
\end{figure}

\begin{figure}[htb]
\centering
\includegraphics[width = \textwidth]{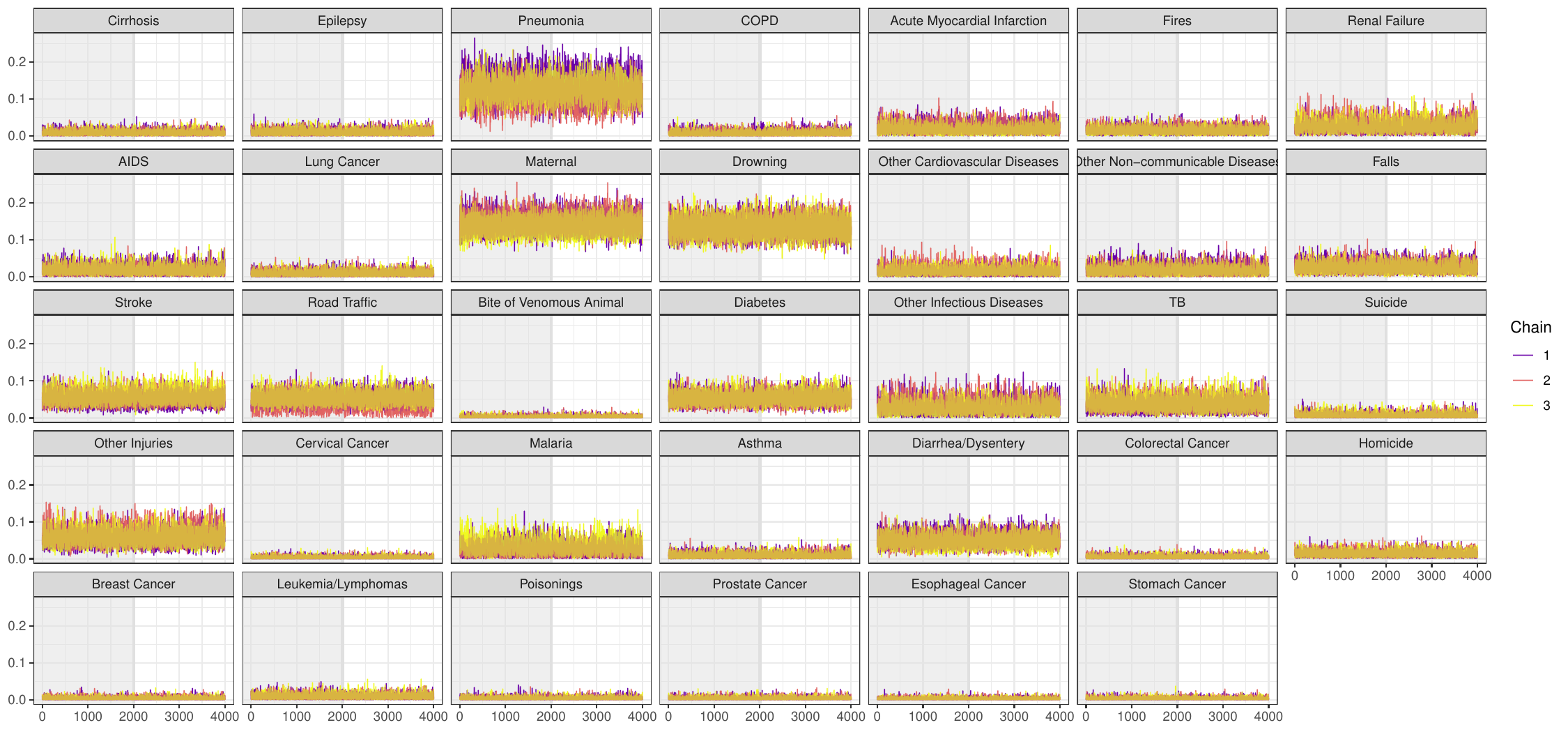}
\includegraphics[width = 0.8\textwidth]{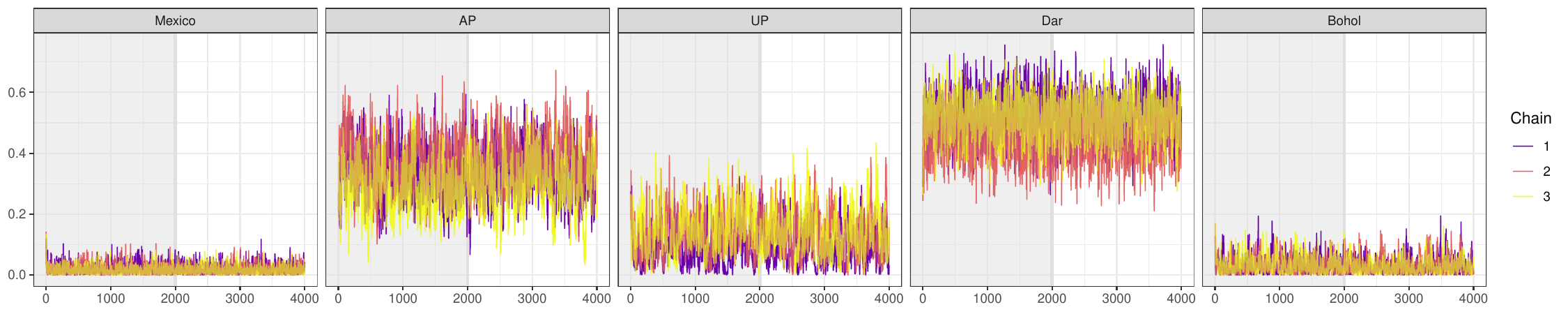}
\caption{Pemba Island as target site: Trace plots for each CSMF, $\pi_c$ (top), and site similarity parameter $\eta_g$ (bottom) from three chains. The burn-in period is shaded in gray.}
\label{Sfig:conv-6}
\end{figure}

In this section, we present additional results of the MCMC convergence analysis. We focus on the multi-domain \nameholder{} model with domain-level shrinkage here. Other variants of the model show similar behavior in the posterior draws.
Figure \ref{Sfig:conv-1} to \ref{Sfig:conv-6} shows the trace plots of the estimated CSMF vector and site similarity parameter $\bm\eta$ from three independent MCMC chains in the prediction stage for each of the six sites in the experiment in Section \ref{sec:results-phmrc-2} of the main paper. Each prediction MCMC chain is fitted with a separate training stage where six training MCMC chains were fitted and stacked to estimate the posterior distribution of the latent class parameters. Overall, the chains mix well and converge to the same distribution across the three independent runs.

\end{document}